\documentclass{aa}  

\usepackage{graphicx}
\usepackage{txfonts}
\usepackage[colorlinks=true,linkcolor=blue, citecolor=blue]{hyperref}
\defcitealias{2020ginolfi}{G20}
\begin{document}

   \title{Simulated [CII] emission in high-$z$ star-forming galaxies}

   \subtitle{}

   \author{N. Mu\~noz-Elgueta\inst{1}\fnmsep\thanks{E-mail: nahir@mpa-garching.mpg.de},
          F. Arrigoni Battaia\inst{1},
          G. Kauffmann\inst{1},
          R. Pakmor\inst{1},
          S. Walch\inst{2,3},
          A. Obreja\inst{4},
          L. Buhlmann\inst{2}
          }

   \institute{Max-Planck-Institut fur Astrophysik, Karl-Schwarzschild-Str 1, D-85748 Garching bei M\"unchen, Germany
              %\email{wuchterl@amok.ast.univie.ac.at}
         \and
             I. Physikalisches Insitut, Universität zu Köln, Zülpicher Str. 77, 50937 Köln, Germany
             %\email{c.ptolemy@hipparch.uheaven.space}
         \and
             Cologne Centre for Data and Simulation Science, University of Cologne, Cologne, Germany
         \and
             Interdisziplinäres Zentrum für Wissenschaftliches Rechnen, Universität Heidelberg, Im Neuenheimer Feld 205, D-69120 Heidelberg, Germany\\
             }

   \date{Received X, X; accepted X, X}

% \abstract{}{}{}{}{} 
% 5 {} token are mandatory
 
  \abstract
   {Extended [CII] emission on tens of kpc, also known as a [CII] halo, is
being currently reported around z$\sim$4-6  star-forming galaxies, especially thanks to the statistics of the ALPINE survey. The [CII]
emission is expected to trace dense cold gas in the inner CGM of these galaxies. The origin of this emission is still debated. In this paper, we present a post-processing model applied to Illustris-TNG50 star-forming galaxies at $z\sim$4-6, and we compare our results with the ALPINE observations. By incorporating C$^{+}$ abundances derived from UV background and young stars as radiation sources, we generate mock observations, from which we extract surface-brightness (SB) profiles. We found that our model predicts similar [CII] emission values on galactic scales as the observations, providing validation for our approach.  However, we find that the predicted [CII] emission in the inner circumgalactic medium (CGM) falls below the observed values by a factor of $\sim$10. We discuss several model limitations that may contribute to this discrepancy. We also find discrepancies with observations when comparing SB profiles of low and high-SFR galaxies. Unlike the observations, simulations exhibit no discernible  difference in the extended [CII] emission between the two subsamples. This discrepancy may reflect shortcomings in feedback model of the simulation. Finally, our analysis suggests that the extended [CII] emission is likely a result of both gas from satellite galaxies and outflows from central galaxies, with satellites playing a dominant role within 0.6$<$R/R$_{\rm vir}<$1. A firm estimate of the importance of each contribution is beyond the scope of the current simulations.}

    \titlerunning{Simulated [CII] in high-z galaxies}
    \authorrunning{N. Mu\~noz-Elgueta et al.}
   \maketitle
%
%-------------------------------------------------------------------

\section{Introduction}
\label{sec:intro}
The gas bound to a galaxy dark-matter halo, %The halo gas surrounding galaxies, 
known as the circumgalactic medium (CGM), plays an essential role in understanding galaxy formation and evolution. It is a reservoir of gas that interacts with the galactic environment, regulating key processes such as star formation, accretion, and feedback. Investigating the CGM provides crucial insights into the complex interplay between galaxies and their surrounding gas reservoirs \citep[e.g.,][]{2017tumlinson}. Studying it is challenging as it is predominantly composed of diffuse and tenuous gas, making direct observations difficult. Recent advancements in observational capabilities \citep[e.g., Atacama Large Millimeter/Submillimeter Array (ALMA),][]{2009wootten}, combined with state-of-the-art simulations (e.g., IllustrisTNG, \citealt{2019nelson,2019pillepich}, EAGLE, \citealt{2015crain,2015schaye}), have opened new avenues for exploring the CGM and unveiling its properties. However, the limited resolution of the simulations and the requirement of deep observations are obstacles that remain.

The [CII] 158$\mu$m fine-structure line has emerged as a valuable tracer of the CGM. This line is one of the brightest emission lines in the infrared spectra of star-forming galaxies \citep[e.g.,][]{1991stacey,2008brauher} and is an important tracer of cold gas in both the interstellar medium (ISM) and CGM. It is emitted from the upper fine-structure level J=3/2 to the lower level J=1/2 of the C$^{+}$ ion. The higher energy level exhibits an equivalent temperature of 91~K, while its critical densities are 16, 2.4$\times 10^{3}$ and 4.8$\times 10^{3}$~cm$^{3}$ for collisions with electrons, hydrogen atoms and H$_{2}$ molecules, respectively \citep{2012goldsmith}. Since the ionization potential of atomic carbon is only 11.3~eV (versus 13.6~eV of neutral hydrogen), [CII] can originate from atomic ISM, diffuse molecular and ionized gas. At high redshifts, both simulations (e.g., \citealt{2015olsen,2015vallini}) and observations (e.g., \citealt{2010stacey,2015gullberg}) have found that the [CII] line primarily emanates from photodissociation regions (PDRs).

Observations of [CII] emission in extragalactic sources notably increased with the advent of the \textit{Herschel Space Observatory} \citep{2010pilbratt}. Studies have focused on local starburst, luminous infrared galaxies, and local dwarf galaxies, investigating the properties of the ISM and its link with the [CII] properties (e.g., \citealt{2012sargsyan,2014diazsantos,2015herreracamus,2017diazsantos,2022samsonyan,2024romano}). In particular, there has been progress in investigating [CII] as a SFR tracer. While different works found a tight correlation between [CII] luminosity and SFR (e.g., \citealt{2014delooze,2015herreracamus}), other \textit{Herschel} observations showed that the $L_{\rm [CII]}$/$L_{\rm IR}$ (or equivalently, $L_{\rm [CII]}$/$SFR$) ratio decreases continuously with increasing $L_{\rm IR}$ (at $L_{\rm IR}\gtrsim$10$^{11}$L$_{\odot}$, e.g., \citealt{2012sargsyan,2017diazsantos,2017contursi}). This could be an indication that the $L_{\rm [CII]}$/$SFR$ ratio is influenced by other galaxy properties, such as metallicity and dust temperature (e.g., \citealt{2013bolatto,2014diazsantos,2023bisbas}).

While locally the [CII] line can only be observed from space due to atmospheric limitations, at high-$z$ the frequency of the [CII] line is 
accessible with ground-based observatories like ALMA.
In particular, the Earth's atmosphere allows us for optimal access to the [CII] transition in the frequency range 345.5-211.2 GHz, which corresponds to the redshift range 4.5$<z<$8.5. %Being easily accessible from the ground for 4.5$<z<$8.5, 
Therefore, [CII] has become a key target for observational studies of high-redshift galaxies, where it can provide insights into the early stages of galaxy formation and evolution. 
Extended [CII] emission (known as [CII] halos) has now been regularly observed in several high-redshift objects such as dusty starburst and normal star-forming galaxies (e.g., \citealt{2019fujimoto,2020fujimoto,2020ginolfi,2020carniani,2020rybak,2015herreracamus,2021herreracamus,2022meyer,2022akins,Solimano2024}). These halos can extend up to tens of kiloparsecs from the galaxies  %(inner CGM at $z>$4), 
and their physical properties and origins are still poorly understood. As reference, the typical effective radius of star-forming galaxies at 4$<z<$6 is expected to be $\lesssim$1.5~kpc (e.g., \citealt{2015shibuya}).
In the context of star-forming galaxies, \cite{2019fujimoto} reported the first identification of [CII] halos ($\sim$10~kpc scales) around $z\sim$5-7 star-forming galaxies using deep ALMA data, through a $uv$-visibility plane stacking. Subsequently, with development of the ALPINE survey \citep{2020lefevre}, further observations of [CII] halos around high-z star-forming galaxies have been reported. 

\cite{2020ginolfi} (hereafter \citetalias{2020ginolfi}) studied a sample of 50 [CII] emitting galaxies at $z\sim$4-6 from ALPINE. By stacking the [CII] spectra and cubes, they find outflows signatures together with extended [CII] emission ($\sim$15~kpc) in their galaxy sample with the higher SFR (SFR$>25$~M$_{\odot}$year$^{-1}$). In particular, they obtained an average [CII] radial  surface-brigthness profile for the high-SFR galaxies, which is characterized by a compact inner component and an extended component (radial scales $>$10~kpc). Later, \cite{2020fujimoto} performed individual measurements for the ALPINE galaxies, finding that $\sim$30~per~cent of them present [CII] halos ($\sim$10 kpc scales). Recently,
an additional ALMA effort, the REBELS survey \citep{2022bouwens} has been conducted. 
This study consists of ALMA [CII] observations of a sample of 40 star-forming galaxies at $z\sim$6.5-9. Based on this survey, \cite{2022fudamoto} stacked 28 [CII] emitting galaxies at $z\sim$7.
Their results indicate that the average [CII] surface-brightness profile is well fit with a single component (i.e., no extended component), with an effective radius of $\sim$2~kpc. Possible reasons of such absence of an extended component could be related to the fact that they stacked moment maps rather than the data cubes, and that they included both high and low-SFR galaxies in the analysis. Despite the aforementioned result, this [CII] emission is spatially more extended compared to the dust continuum and the rest-frame UV emission, in agreement with previous works (e.g., \citetalias{2020ginolfi}, \citealt{2020fujimoto}).

To fully understand these [CII] halos, it is essential to perform theoretical investigations. Currently, cosmological simulations are not able to reproduce extended [CII] emission at high-$z$ (e.g., \citealt{2017pallottini,2019pallottini}). For instance, \cite{2019fujimoto} compared their observational results with two sets of numerical simulations of massive galaxies ($M_{\rm halo}\sim$10$^{11}$-10$^{12}$M$_{\odot}$) at $z\sim6$ (\citealt{2017pallottini,2019arata}, $m_{\rm baryon}$ = 1.2$\times$10$^{4}$ and 1.8$\times$10$^{5}$~M$_{\odot}h^{-1}$, respectively). By using mock data from both simulations, they find that the [CII] emission is not as extended as the observational data. Specifically, they obtained surface-brightness profiles which decline rapidly beyond $r\sim$3~kpc, which is significantly smaller than the reported [CII] extension of 10~kpc observed by \cite{2019fujimoto}.
These profiles are at least one order of magnitude lower than the observed flux at inner-CGM scales. This disagreement suggests that certain crucial physical processes for the production of [CII], such as feedback or metal enrichment,  are not well reproduced by the simulations. It is also possible that there are limitations or uncertainties in the modelling of [CII] emissivities.
This discrepancy between simulations and observations highlights the need for additional investigations and theoretical refinement.

The origin of this extended [CII] emission remains unclear. Different scenarios have been proposed, including the presence of satellite galaxies, extended PDRs, extended HII regions, inflows of cold streams, outflows produced by stellar feedback or AGN feedback (see \citealt{2019fujimoto} for a detailed description). 
Currently, the scenario in which star-formation driven outflows (from the central galaxy) are responsible for the extended [CII] emission %scenario 
seems to be the most plausible according to observations (e.g., \citetalias{2020ginolfi}, \citealt{2020fujimoto, 2024romano}). Indeed the observed [CII] halos are only evident around more star-forming galaxies (SFR$>25$~M$_{\odot}$year$^{-1}$) and the line-emission stacking of these sources seems to highlight the presence of broad wings \citepalias{2020ginolfi}. This scenario has also been supported by semi-analytical models (e.g., \citealt{2020pizzati,2023pizzati}). These models primarily focus on simulating the cooling outflows and their resulting [CII] emission, suggesting that the observed [CII] halos could be attributed to gas that is cooling after being heated and expelled in supernova-driven outflows. The effectiveness of these models is largely attributed to their ability to accurately capture the catastrophic cooling 
process within central kpc scales. Specifically, they find that supernova-driven outflows expand into the CGM at velocities of 200-500 km~s$^{-1}$. The outflowing gas undergoes rapid cooling ($\sim$few hundred~K) within the first kpc, then it is slowly heated (T$\sim$10$^{3}$~K) by the cosmic UV background. These conditions favor the formation and survival of C$^{+}$ ions. 

In this framework, we post-processed 
a sample of Illustris-TNG50 halos hosting z$\sim$4-6 star-forming galaxies to model their [CII] emission.
In particular, we produce mock observations and compare them with the ALPINE observations \citepalias{2020ginolfi}, ultimately discussing the possible origin for the observed extended [CII] emission in the inner CGM of high-z galaxies. This paper is structured as follows. In section \ref{sec:observations} we present a summary of the observations of \citetalias{2020ginolfi}. In section \ref{sec:methods} we describe the methods used in this work. In section \ref{sec:results} we present our main results. In section \ref{sec:discussion} we discuss our main results, explore the different factors involved in the production of [CII] halos and caveats. Finally, section \ref{sec:summary} summarizes our findings.

\section{OBSERVATIONS} % added by FAB
\label{sec:observations}

The ALPINE survey is an ALMA large program that targets [CII] and FIR-continuum emission of a sample of 118 normal (i.e., main sequence) star-forming galaxies. These galaxies are at 4.4$<z<$5.8, and have M$_{*}\sim$10$^{8.5}$- 10$^{11}$~M$_{\odot}$ and SFR$\sim$3-650~M$_{\odot}$year$^{-1}$. The average resolution of these observations is 0.9~arcsec, i.e., ranging from $\sim$5.2 to 6~kpc for the different redshifts considered. This coarse resolution implies that the galaxies are not resolved in these ALMA observations. The spectral resolution is $\Delta \nu_{\rm channel}$ = 25 - 35 km~s$^{-1}$, enabling the resolution of the typical [CII] emission line width ($FWHM\sim$~235~km~s$^{-1}$). All the details regarding the observational setup and data reduction are described in \cite{2020bethermin}.
In total, 75 out of 118 galaxies were robustly detected in [CII], with signal-to-noise ratio (S/N) $>$3.5. \citetalias{2020ginolfi} studied these [CII]-emitting galaxies, excluding $\sim$30$\%$ of those objects that present signs of ongoing mergers (major or minor). Thus, their final sample\footnote{Some of the physical properties of Ginolfi's sample listed here are different than %might be inconsistent with 
the values reported in \citetalias{2020ginolfi}. These discrepancies are due to the fact that \citetalias{2020ginolfi} used the values from a preliminary analysis of the ALPINE sample. Here, we use 
%arises because we have used %included 
the values reported in the catalogs of the ALPINE survey \citep{2020lefevre}, which was published after \citetalias{2020ginolfi}.} consists of 50 normal star-forming galaxies at 4.4$<z<$5.8, with M$_{*}\sim$10$^{8.5}$- 10$^{11}$~M$_{\odot}$ and SFR$\sim$5-650~M$_{\odot}$year$^{-1}$.

In order to investigate the efficiency of galactic feedback in these galaxies and the spatial distribution of the [CII] emission, \citetalias{2020ginolfi} investigated the [CII] emission data through a stacking analysis. For the purposes of this paper, in the following paragraphs we summarize only their stacking of the cubes.
This procedure consists of a combination of the [CII] cubes of each galaxy C$_{i}$ with $i$ slices (previously aligned), based on a vector variance stacking technique, where the stacked cube C$_{i}^{\rm stack}$ is estimated as:
\begin{equation*}
    C_{i}^{\rm stack} =   \dfrac{\sum_{k=1}^{N} C_{i,k} \cdot w_{i,k}}{\sum_{k=1}^{N} w_{i,k}} 
\end{equation*}

where C$_{i,k}$ corresponds to the cube of the $k$th galaxy. The weigthing factor w$_{i,k}$ is the inverse variance, equal to 1/$\sigma^{2}_{i,k}$, where $\sigma_{i,k}$ corresponds to the spatial rms associated with each slice of each galaxy. In this way, noisier cubes are down-weighted.

This stacking was performed %for the full sample and also 
%separately for 
on two sub-samples separated by a threshold in SFR: high-SFR (SFR $>$ 25~M$_{\odot}$~yr$^{-1}$) galaxies and low-SFR (SFR $\leq$ 25~M$_{\odot}$~yr$^{-1}$) galaxies. The SFRs were computed through spectral energy distribution (SED) fitting, assuming a constant star-formation history over 100~Myr \citep{2020faisst}.
In order to obtain velocity-integrated flux maps of the core of the [CII] emission, the spectral slices of the stacked cubes were collapsed in the velocity range of [-200: +200]~km~s$^{-1}$.
The final stacked synthetized beam (or point spread function, PSF) was estimated by stacking the PSF of each input galaxy cube. Then, the channels of this stacked PSF cube are collapsed over the same velocity range mentioned above. This final PSF has a  major axis full-width-half-maximum (FWHM) of 0.98 arcsec, a minor axis FWHM of 0.89~arcsec, and a position angle of -30 deg.
From the resulting [CII] flux maps of the subsamples, \citetalias{2020ginolfi} finally computed circularly averaged radial surface-brigthness (SB) profiles, considering radial bins of 0.15~arcsec. 
For most of our comparative analysis, we present a single observed SB profile, derived by averaging the two SB profiles obtained from \citetalias{2020ginolfi} for the high and low-SFR galaxy subsamples.

\section{Methods}
\label{sec:methods}
%observation, simulations
\subsection{IllustrisTNG simulations}
\label{sec:illustristng}

In this work, we explore different high-$z$ halos from the IllustrisTNG cosmological magnetohydrodynamical simulations \citep{2019nelson, 2019pillepich}. Using the moving-mesh code AREPO \citep{2010springel}, these simulations follow the coupled dynamics of the dark-matter and gas. This method is based on a spatial discretization consisting of a Voronoi tessellation of the simulation box. The physical processes of baryonic TNG runs, among which are star-formation, stellar evolution and feedback, supermassive black hole formation and evolution, follow the ``TNG galaxy formation model'' \citep[methods described in][]{2017weinberger,2018pillepich}. We briefly describe here only some aspects of these simulations relevant for this work.

The simulations include gas radiative processes, including metal-line cooling and heating from the ultraviolet background (UVB). Due to limitations in the resolution, the gas does not cool radiatively below 10$^{4}$~K and the processes of star formation and pressurization in the multiphase ISM are modeled based on the approach of \cite{2003springel}. In this model, gas with densities above a threshold of $\simeq$~0.1~cm$^{-3}$ undergoes stochastic star formation according to the Kennicutt-Schmidt relation, %\footnote{The Kennicutt-Schmidt law is an empirical relation between the SFR surface density and the gas surface density in galaxies.}, 
assuming a Chabrier initial mass function \citep{2003chabrier}.
To account for the effects of unresolved supernovae, a two-phase effective equation of state model is employed to incorporate the pressurization of star-forming gas. Chemical enrichment is driven from supernovae type II, Ia, and AGB stars. The simulations also include local radiation effects from AGN, and AGN feedback in two modes: a thermal ‘quasar’ state at high accretion
rates, and a kinetic ‘wind’ state at low accretion rates (\citealt{2017weinberger}).

This TNG model has been calibrated in order to reproduce several of the observed galaxy properties mainly at $z$=0. The cosmology assumed 
%is according the \cite{2016planck}, with 
has parameters $\Omega_{\Lambda,0}$= 0.6911 , $\Omega_{m,0}$ = 0.3089, $\Omega_{b,0}$ = 0.0486, $\sigma_{8}$ = 0.8159, $n_{s}$ =  0.9667 and $h$ = 0.6774 (\citep{2016planck}).

Specifically, in this work we used the TNG50-1 box, which presents the smallest volume (51.7$^{3}$~cMpc$^{3}$) but the highest resolution ($m_{\rm baryon}$ = 5.6$\times$10$^{4}$~M$_{\odot}h^{-1}$, galaxies with M$_{*}\gtrsim$10$^{7}$~M$_{\odot}$) of the TNG runs. The latter is a key factor to perform our [CII] emission predictions, especially considering that high densities are needed for [CII] production (see critical densities in Section~\ref{sec:intro}).
Compared to previous works, the TNG50 simulations offer key advantages. They strike a balance between computational efficiency and resolution, allowing for a larger sample size and a more comprehensive representation of the high-$z$ halo population. Additionally, these simulations incorporate a self-consistent treatment of various physical processes, which are uniformly implemented across the simulation volume. This ensures a consistent modeling approach for the entire sample of halos.

\subsection{Selection of simulated galaxies}
\label{sec:selection_sample}

We selected a sample of TNG50-1 star-forming galaxies\footnote{In the TNG simulation suite, dark-matter haloes are identified using a Friends-of-Friends (FoF) algorithm \citep{1985davis}, and subhalos are identified as gravitationally bound substructures through the \textsc{subfind} algorithm \citep{2005springel}. In this work, the selected TNG50-1 galaxies consist in the central subhalos of their corresponding FoF halos.} 
with similar physical properties to the galaxies of the full ALPINE survey \citep{2020lefevre}. Specifically, they were chosen to match the ranges in stellar mass (log(M$_{*}$) $\sim$8.5 - 11.0~M$_{\odot}$) and star-formation rate (SFR $\sim$3 - 623 M$_{\odot}$year$^{-1}$)\footnote{About 37~per~cent of the TNG50-1 galaxies with M$_{*}$ in the range $\sim$8.5 - 11.0~M$_{\odot}$ have SFRs in the range $\sim$3 - 623 M$_{\odot}$year$^{-1}$.}. In total, we found 357 TNG50 galaxies matching these properties. We then optimized our selection by constraining our distribution of SFR and M$_{*}$ to conform to those exhibited by the ALPINE [CII] emitting galaxies \citepalias{2020ginolfi}.
Throughout this work, we adopt for the simulated galaxies:  
i) stellar masses within twice the stellar half mass radius, and ii) SFRs averaged across the last 100 Myrs within twice the stellar half mass radius \citep[see][]{2019donnari, 2019pillepich}, comparable to the observationally derived SFRs %based on the UV luminosity at 2300~\AA\ 
(see Section~\ref{sec:observations}). These galaxies were also selected to have similar redshifts to those of the observations ($z\sim$4.4-5.7). However, it was not possible to precisely match them because snapshots with intermediate redshifts (the so-called mini snapshots) do not contain information on the metal abundances which are needed to estimate the [CII] emission. Thus, the final selected halos were extracted
from the three full snapshots (13, 17 and 21) corresponding to the redshifts 6, 5 and 4, respectively. 

\begin{figure}
    \centering
    \includegraphics[width=0.9\columnwidth]{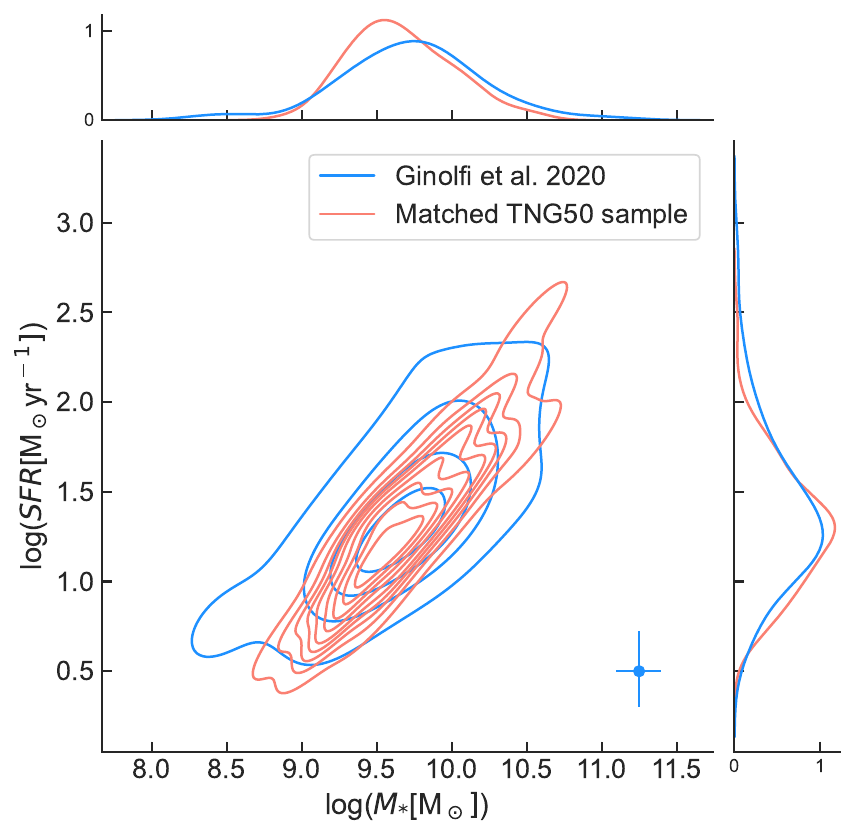}
    \caption{Comparison of the distribution of the SFRs and M$_{*}$ values between the simulated (red) and observed (blue) samples (75 and 50 galaxies, respectively). The joint distribution is visualized using density contours. The contour spacing represents the density of data points, with closer contours indicating regions of higher data density. The blue bars in the bottom-right corner represent the average 1-sigma error for the SFR and M$_{*}$ of the observations. The histograms along the x and y axes provide marginal distributions of the M$_{*}$ and SFR, respectively, for both observed and simulated samples.}
    \label{fig:hist_sample}
\end{figure}

The individual central galaxies were cut from the full snapshot (at $\sim$3~R$_{200}$) in order to facilitate their analysis. In total, we selected 75 galaxies that match the distribution of SFR and M$_{*}$ of the sample in \citetalias{2020ginolfi}. Analogous to the observations, we excluded from our analysis those systems that would have exhibited observable merging signatures. Specifically, following the observational limitations, we did not consider
systems with companion galaxies with SFR$>$1.5~M$_{\odot}$year$^{-1}$, within a 3D radial distance of 1.1R$_{200}$. 
We also note that our selected galaxies contain AGNs, while the observed sample excludes type I AGNs. However, these AGNs of our simulated sample have low bolometric luminosities\footnote{We estimated the bolometric luminosity of the central AGNs following \cite{2019nelson}.} (L$_{\rm BOL}\lesssim$10$^{45.5}$~erg~s$^{-1}$), which would not be detected in currently available observations. 
Figure \ref{fig:hist_sample} shows the distribution of the selected simulated galaxies (red) in the SFR and M$_{*}$ plane, in comparison to the sample of \citetalias{2020ginolfi} (blue), using density countours. Importantly, this selection represents one possible realization of the sample in \citetalias{2020ginolfi} extracted from the TNG50 galaxies in the mass and SFR ranges of the ALPINE survey. This means that there could be other selections of simulated galaxies that also fit the same criteria of the sample in \citetalias{2020ginolfi}, but include slightly different individual galaxies. In this figure, the spacing between the blue contours (observations) appears larger than the spacing between the red contours (simulations). This suggests that the observed sample has a more spread-out distribution of galaxies in the joint space of M$_{*}$ and SFR compared to the simulated sample. Areas where the blue and red contours overlap suggest regions of similar M$_{*}$ and SFR values between both samples.

\subsection{[CII] cooling rate and flux estimation}\label{sec:cooling_rates}

In order to obtain an estimate of the [CII] emission for each selected halo, we used the analytical [CII] cooling rate ($\Lambda_{\rm CII}$) from \cite{2022bisbas}. %, which is applicable to $T>$150~K. 
In their study, \cite{2022bisbas} performed [CII] synthethic observations using the \textsc{griffin} smooth particle hydrodynamics simulations \citep{2020lahen}. These simulations have a resolution of 4M$_{\odot}$ per gas particle and use a modified version of \textsc{gadget-3} \citep{2005springelb} described in \cite{2014hu}. They assume a non-equilibrium model of cooling and chemistry that traces H$_{2}$, H$^{+}$, CO, H, C$^{+}$, O and free electrons (\citealt{1997nelson,2007glover}), only when the gas temperature is $<$3$\times$10$^{4}$~K. For temperatures above this threshold, equilibrium cooling tables based on 12 different metal species (H, He, N, C, O, Si, Mg, Fe, S, Ca, Ne, and Zn) are used (from \citealt{2009wiersma}). Furthermore, the simulations also include routines to account for the interstellar UV radiation field and mechanisms for stellar feedback, such as photoionization, photo-electric heating, and supernovae. 

Based on the direct $\Lambda_{\rm CII}$ estimations from the simulations, \cite{2022bisbas} proposed a theoretical approach to get $\Lambda_{\rm CII}$, based on analytical expectations. 
They demonstrated that the $\Lambda_{\rm CII}$ values obtained from the analytical expressions are in agreement with the majority of the $\Lambda_{\rm CII}$ values obtained directly from their simulations.
This approach assumes that the gas is optically thin and is applicable for $T>$150~K. The analytical cooling rate can be expressed as:\\ %\citep{2012goldsmith}

\begin{equation}
\label{eq:lambda}
    \Lambda_{\rm CII} = \dfrac{R_{\rm c,dex}\times 2e^{-91.25/\rm T}}{R_{\rm s}+R_{\rm c,dex}(1+2e^{-91.25/\rm T})}  R_{\rm s}\ E_{\rm CII}\ n_{\rm C^{+}} (\rm erg\ s^{-1}\ cm^{-3}),
\end{equation}
where $T$ is the temperature of each gas cell and the excitation temperature of CII transition is 91.25~K, E$_{\rm CII}$ is the energy $h\nu_{\rm CII} = 1.25988 \times 10^{-14}\, \rm erg$, $n_{\rm C^{+}}$ is the number density of C$^{+}$ particles, R$_{\rm s}$ is the spontaneous emission equal to the corresponding Einstein $A$-coefficient R$_{\rm s}$=2.291$\times$10$^{-6}$s$^{-1}$, and R$_{\rm c,dex}$ is the total de-excitation rate given by: 
\begin{equation}
\label{eq:rates}
    R_{\rm c,dex} = R_{\rm c,10}({\rm H}) n_{\rm H} + R_{\rm c,10}({\rm H_{2}})n_{\rm H_{2}}  + R_{\rm c,10}(e-)n_{\rm e}, 
\end{equation}

\noindent where R$_{\rm c,10}$(e-), R$_{\rm c,10}$(H$_{2}$) and R$_{\rm 10}$(H) are  collisional de-excitation rates with e-, H,
and H$_{2}$ as colliding partners, respectively. We consider the contribution of collisions with H$_{2}$ negligible. Indeed, previous works argue that the contribution of molecular gas to the total [CII] luminosity is minor (e.g., \citealt{2018franeck,2021tarantino,2022bisbas}). Therefore, we focus on [CII] emission originating from PDRs where sufficient amounts of HI and e- are present as collision partners (e.g., \citealt{2015olsen}). This assumption is also based on the fact that TNG50 does not follow the formation and evolution of molecular gas, consistent with the simulation setup.

Then, R$_{\rm c,10}$(e-) and  R$_{\rm c,10}$(H) are (from Leiden Atomic and Molecular Database, \citealt{2005schoeier}):

\begin{equation}
    R_{\rm c,10}(\rm e-) = 2.426206 \times 10^{-7} \left(\dfrac{T}{100}\right)^{-0.345} (\rm cm^{-3}\ s^{-1})
\end{equation}
\begin{equation}
    R_{\rm c,10}(\rm H) = 3.113619 \times 10^{-10} \left( \dfrac{T}{100}\right)^{0.385} (\rm cm^{-3}\ s^{-1})
\end{equation}

To estimate $\Lambda_{\rm CII}$ (equation~\ref{eq:lambda}), all required parameters except $n_{\rm C^{+}}$ can be estimated or assumed from the outputs of TNG50-1 (i.e., temperatures and densities).

\begin{figure}
    \centering
    \includegraphics[width=\columnwidth]{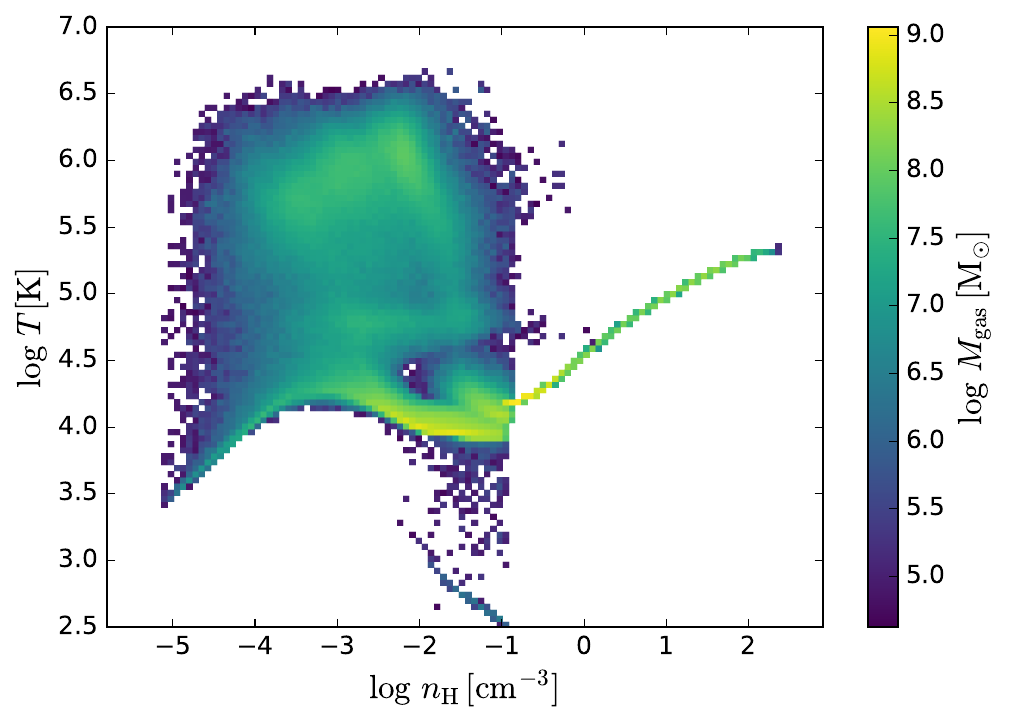}
    \caption[Phase-diagram for one of the selected TNG50 galaxies at $z$=5.]{Phase-diagram for one of the selected TNG50 galaxies at $z$=5, weighted and colour coded by gas mass. The diagram encompasses all gas cells within R$_{200}$. For densities above the star-formation threshold $n_{\rm H}$=0.1, the temperatures are determined by the effective equation of state \citep{2003springel}.}
    \label{fig:phase}
\end{figure}

To gain insight into the physical properties of the selected galaxies, Figure \ref{fig:phase} presents a temperature-density ($T - n_{\rm H}$) phase diagram within R$_{200}$, of one $z$=5 system. The color bar indicates the gas mass distribution, where the yellow (blue) regions correspond to areas of high (low) gas mass concentration within the halo. The region where most of the mass is concentrated ($\sim T<$10$^{4.5}$~K, 10$^{-2.5}<n_{\rm H}<$10$^{-1}$cm$^{-3}$) is dominated by ISM gas. Note that all dense gas cells with $n_{\rm H}>$0.1~cm$^{-3}$ are by construction star-forming in these simulations. The temperatures of this star-forming gas is determined by the effective equation of state \citep{2003springel} and they do not represent the physical temperatures of the gas \citep{2019pillepich}. Instead, the star-forming gas cells ($n_{\rm H}>0.13$~cm$^{-3}$) are assumed to be at their cold-phase temperature of 1000~K \citep{2003springel}, as the cold-phase is dominant in mass with respect to the warmer component in this regime. This approach is common in modelling emission lines arising from the cold-phase in TNG simulations (e.g., \citealt{2021nelson}).

\begin{figure}
    \centering
    \includegraphics[width=\columnwidth]{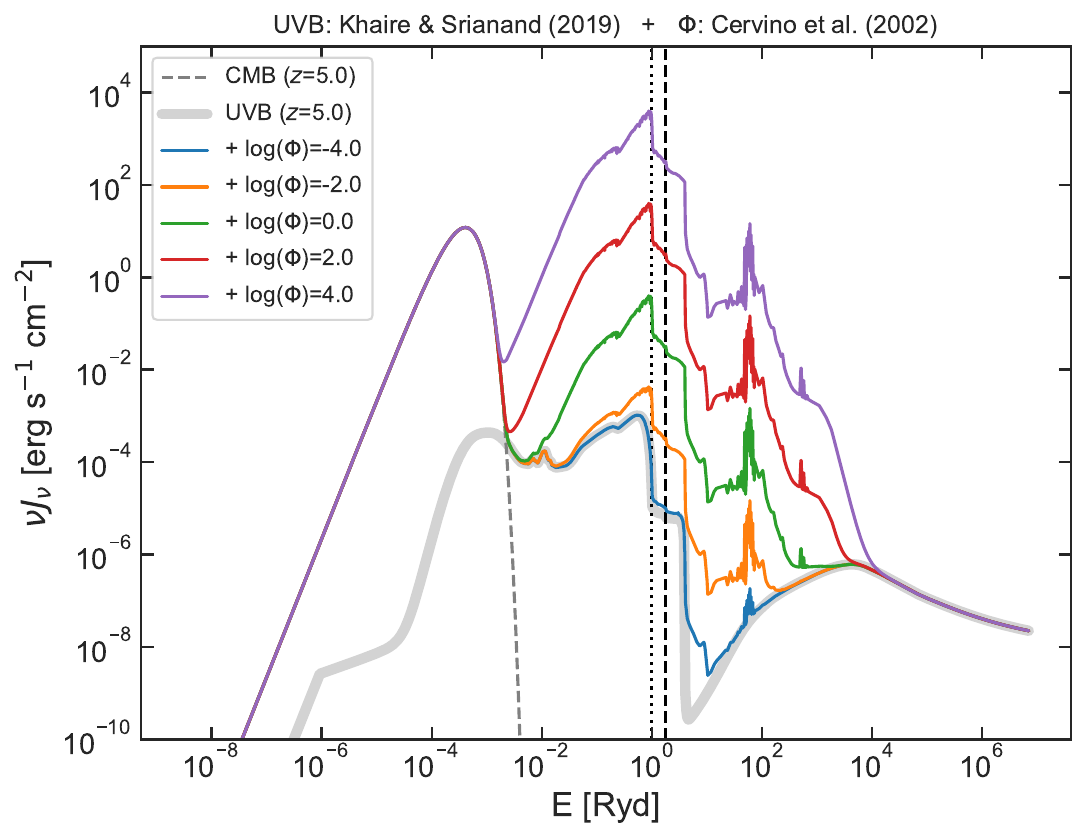}
    \caption[Example of different input spectra used in the \textsc{cloudy} calculations.]{Example of different input spectra used in the \textsc{cloudy} calculations. The solid light-grey line represents the UVB SED of \cite{2019khaire}, and the dashed grey line the CMB, both at $z$=5. The vertical black lines represent the ionization energies of C$\rightarrow$C$^{+}$ (dotted) and C$^{+}$$\rightarrow$C$^{++}$ (dashed). The coloured lines represent the contribution to the SED due to  young stars \citep{2002cervino} at different normalizations, plus the contribution of the UVB. The units of the radiation field $\phi$ are $\rm M_{\odot}\ year^{-1}\ kpc^{-2}$. }
    \label{fig:input_spectra}
\end{figure}

In order to estimate the values of n$_{\rm C^{+}}$ we need the C$^{+}$ abundances for each gas cell. As an initial simplified approach, we assume a scenario where all (i.e., the total) neutral carbon is converted into C$^{+}$. Subsequently, we explore other cases in which the C$^{+}$ abundances are obtained using the spectral synthesis code \textsc{cloudy} version 17.03 \citep{2017ferland}. For the \textsc{cloudy} calculations, we either assumed the redshift dependent UVB of \cite{2019khaire} as the only photoionization source, or the combination of UVB and radiation from nearby young stars (ages $<$10 Myr). For the young stars we used the spectra model of \cite{2002cervino}, and assumed that only a fixed 5~per~cent of the stellar radiation can escape the molecular birth cloud (i.e., escape fraction $f_{\rm esc}$ = 0.05). The normalization of the stellar radiation field ($\phi$) in units of $\rm M_{\odot}\ year^{-1}\ kpc^{-2}$ at each gas particle position $\overrightarrow{r}_j$ was computed in the optical thin approximation (no absorbtion along the line-of-sight beyond the constant at-the-source $f_{\rm esc}$): 
\begin{equation}
    \phi(\overrightarrow{r}_j) = \frac{f_{\rm esc}}{10^7yr}\sum_i \frac{M_i}{(\overrightarrow{r}_i-\overrightarrow{r}_j)^2},
\end{equation}
where $M_i$ and $\overrightarrow{r}_i$ are the mass and position of stellar particles with ages less than 10~Myr, following \citet{2019obreja}. To compute the ionization fraction $\chi=n_{\rm C^+}/n_{\rm C}$, we thus run one zone plane-parallel \textsc{cloudy} models for fixed hydrogen number density log($n_{\rm H}$), temperature log($T$), metallicity log($Z$/$Z_{\rm\odot}$), normalization of the young star radiation field log($\phi$) and redshift ($z=3,4,5$ for the UVB), to cover all the physical conditions of gas cells in the TNG50 galaxies. No dust and hence no metal depletion is assumed in accordance with the TNG50 simulation setup. Subsequently, we used 3D or 4D barycentric interpolation in either $\chi$[log($Z$),log($n_{\rm H}$),log($T$)] or $\chi$[log($\phi$,log($Z$),log($n_{\rm H}$),log($T$)] \textsc{cloudy} tables, to get the C$^+$ ionization fraction of each gas cell for the UVB and UVB plus young star photoionization models at fixed redshift. Figure \ref{fig:input_spectra} shows examples of input spectra used in the \textsc{cloudy} calculations, and Appendix~\ref{sec:appendix2} gives an example of a \textsc{cloudy} input file.

Once the C$^{+}$ abundances are computed, all the quantities in equation~\ref{eq:lambda} are available.
We can then obtain the [CII] luminosities per cell as  $L_{\rm [C{\sc II}]} = \Lambda_{\rm CII} \times V $, with $V$ the volume of each cell, 
and estimate their respective flux densities following \cite{1992solomon}:

\begin{equation}\label{eq:luminosity}
    I_{\rm [C{\sc II}]} = \dfrac{L_{\rm [C{\sc II}]}}{1.04\times 10^{-3} \nu_{\rm obs} D_{\rm L}^{2} } (L_{\odot}) (\rm Jy~km~s^{-1}),
\end{equation}
where $L_{\rm [C{\sc II}]}$ is the [C{\sc ii}] line luminosity in L$_{\odot}$ units, $\nu_{\rm obs}$ is the observed frequency in units of GHz, and $D_{\rm L}$ corresponds to the luminosity distance in Mpc.

\subsection{[CII] 2D flux maps and surface brightness profiles}
\label{sec:analysis_profiles}
After getting $L_{\rm [C{\sc II}]}$ for each gas cell, we constructed %proceed to perform 
[CII] surface brightness maps. For this, we obtained 2D projections by integrating the emission within a slice encompassing a range of [-R$_{200}$, +R$_{200}$] (or $\sim$[-40, +40]~kpc at the median $z$=5) from the galaxy's center\footnote{The center of the galaxies considered here is the position of the particle with the minimum gravitational potential energy.}. As a conservative approach, we did not apply any cut in velocities, ensuring the inclusion of the central galaxy in our analysis. Beyond the chosen range, structures at larger distances are not expected to greatly contribute to the [CII] signal, which is dominated by the central galaxy. In contrast, the observed maps apply a velocity cut of [-200,+200] km~s$^{-1}$.
For reference, these velocities are similar to the virial velocity for the average halo hosting the selected TNG50 galaxies ($v_{\rm vir}\sim220$~km~s$^{-1}$).

\begin{figure*}
    \centering
    \includegraphics[width=0.8\textwidth]{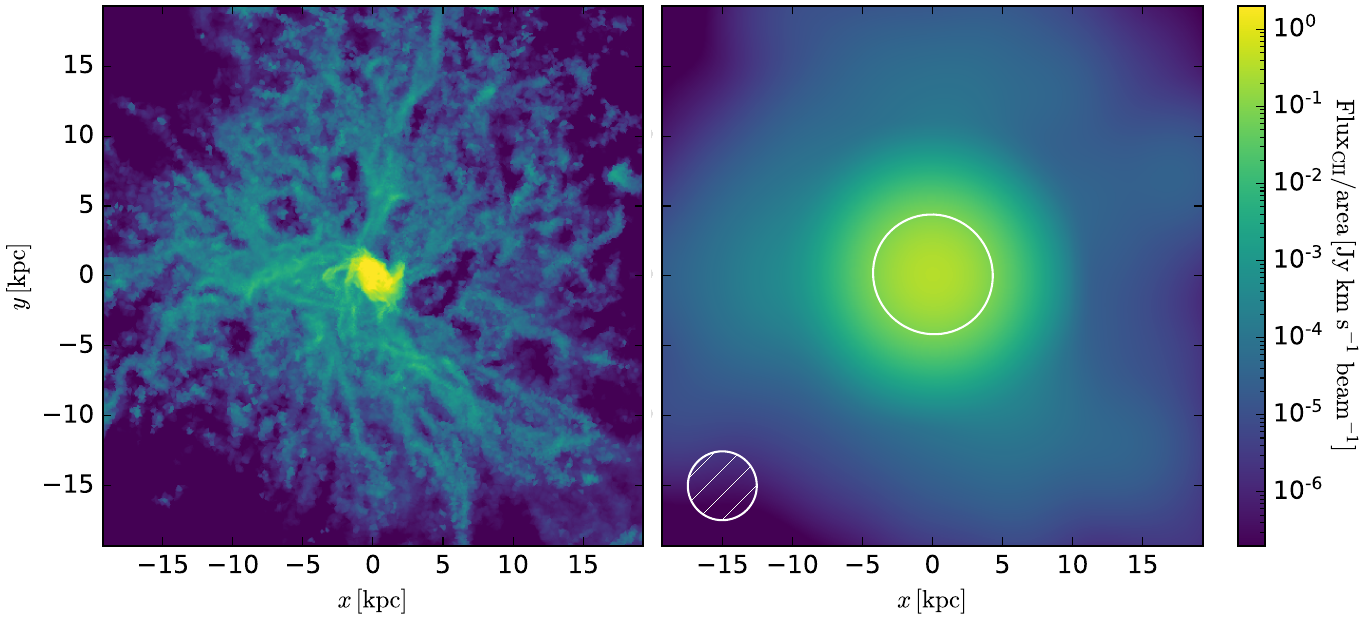}
    \caption[{[CII]} emission surface-brightnes map for one of the selected TNG50 galaxies at $z$=5.]{[CII] emission surface-brightnes map for one of the selected TNG50 galaxies at $z$=5, assuming C$^{+}$ abundances with UVB and local young stars as ionization sources. Left: the map is presented at the full-resolution of the simulation. Right: the map has been convolved with the average ALMA beam of the \citetalias{2020ginolfi} observations (bottom left corner). The white contour represents the average 2$\sigma$ level of the observations (0.08~$\rm Jy\ km\ s^{-1}\ beam^{-1}$). }
    \label{fig:halo_2dflux}
\end{figure*}

In order to make these 2D maps comparable with the observations, we applied a 2D Gaussian smoothing with a FWHM equal to the stacked synthetized beam obtained in the ALMA observations (0.9~arcsec or $\sim$6~kpc at $z$=5, see Section \ref{sec:observations}).
Figure \ref{fig:halo_2dflux} shows an example of [CII] 2D surface brightness %flux 
maps for one of the selected galaxies at $z$=5, at full resolution\footnote{The spatial resolution in the simulated halos varies with density or radius from the center of the halo. For reference, the luminosity weighted average cell radius at the center, 1 arcsec (or 6.3~kpc) and 5 arcsec (or 31.5~kpc) in and around a TNG50 galaxy at z=5 is 50~pc, 0.3~kpc and 1~kpc, respectively.} (left panel) and with smoothing (right panel). We have not added noise to our modeled smoothed SB maps because our primary focus is to assess whether we can achieve similar flux levels as observed. Future simulations of ALMA observations with \textsc{simalma}\footnote{\textsc{simalma} is a task from CASA used to simulate observations with ALMA and the Atacama Compact Array (ACA).} are needed to incorporate noise levels and achieve a more realistic representation.

For each galaxy, we then extract  circularly averaged radial profiles from their 2D smoothed SB maps. The profiles were obtained by considering radial bins of 0.15~arcsec, as done in the observations. Subsequently, total averaged profiles were obtained by stacking the individual profiles. To accomplish this, we utilized a total of 20 orientations per galaxy and performed 1000 iterations of the experiment, randomly selecting one orientation per galaxy and calculating the final average based on these realizations. The range covered by the resulting profiles from these different orientations is indicated by the shaded areas in Figure \ref{fig:sb_profiles}.

\section{Results}
\label{sec:results}
In this section, we present the results obtained from our post-processing model, which generates [CII] emission in high-redshift galaxies using the TNG50 simulations. By comparing our model predictions with observational data from the ALPINE survey, we assess the validity and performance of our approach, and try to provide some intuition on the aforementioned [CII] emission scenarios.

\subsection{Relation between [CII] luminosity and star formation rate}
\label{sec:cii_sfr_relation}
Several studies have found that the [CII] luminosity is tightly correlated with the SFR, with a nearly linear relation \citep[e.g,][]{2014delooze,2017olsen,2018lagache, 2020schaerer}. 
However, at high-$z$ the scatter around the local relation tends to increase. This fact has been primarily attributed to variations in galaxy ISM physical properties, such as gas density and metallicity (e.g., \citealt{2018lagache}). %The same relation but with larger scatter has been found for high-z galaxies, mainly because differences in their physical properties (i.e., gas density, metallicity). 
In this context, we explore the [CII]-SFR relation obtained from our post-processing model, and compare the obtained results with the observational data from the ALPINE survey. Therefore, the subsequent analysis  checks the consistency between our model predictions and the observations.

\begin{figure*}
    \centering
    \includegraphics[scale=0.56]{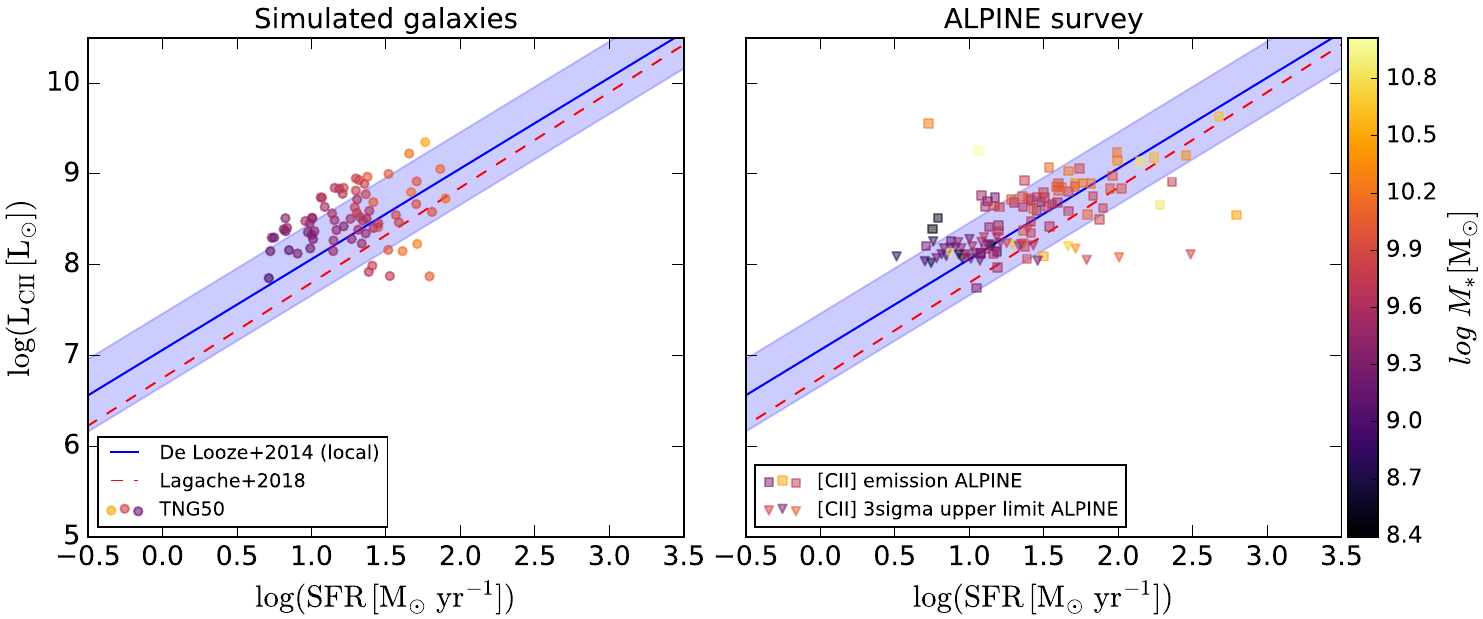}
    \caption[{[CII]} luminosity versus SFR for the TNG50 galaxies and the ALPINE survey.]{[CII] luminosity versus SFR for the TNG50 galaxies (left panel) and the ALPINE survey (right panel). The shaded blue area represents the 1$\sigma$ relationship of \cite{2014delooze} for starburst local galaxies. The dashed red line shows the relationship of \cite{2018lagache} for $z$=5 star-forming galaxies. %The black box in the left panel contains three simulated galaxies with L$_{[\rm CII]}$ below the observational limit of the ALPINE survey. 
    The average 1$\sigma$ error for L$_{[\rm CII]}$ and SFR in the right panel are 0.05~dex and 0.4~dex, respectively.}
    \label{fig:lum_sfr}
\end{figure*}

The results are shown in Figure \ref{fig:lum_sfr}, where both the simulated (left panel) and observational (right panel) data are color-coded by stellar mass. For the simulated galaxies, the x-axis shows the SFR averaged over the last 100 Myrs, and the y-axis the [CII] luminosity obtained by assuming UVB and young stars as photoionization fields, and by integrating inside the area within the 2$\sigma$ level of the observations (stacked flux map, r$\sim$1.3~arcsec). Comparing the two panels, it is evident that our results are in agreement with the observations, when considering the simulated galaxies that can be detected by the observations ($L_{\rm [CII]} > $10$^{7.8} \rm L_{\odot}$). In addition, for both sets of data we can appreciate a trend in the stellar masses along the relation. However, when comparing the distributions of normalized distances\footnote{The normalized Euclidean distances for the ALPINE and TNG50 datasets are estimated as: \begin{multline*}
    r_{\text{data}} = \left(\frac{{(M_{\rm *data} - M_{\rm *min})^2}}{{(M_{\rm *max} - M_{\rm *min})^2}}\right) + \left(\frac{{(SFR_{\rm data} - SFR_{\rm min})^2}}{{(SFR_{\rm max} - SFR_{\rm min})^2}}\right)\\
    + \left(\frac{{(L_{\rm[CII] data} - L_{\rm[CII] min})^2}}{{(L_{\rm[CII] max} - L_{\rm[CII] min})^2}}\right)
    \end{multline*}
where the subscript $data$ refers to the specific data set being considered (ALPINE or TNG50), $min$ and $max$ represent the minimum and maximum values obtained for the variables $M_{*}$, SFR and $L_{\rm [CII]}$ between the two samples, respectively. 
} 
between $M_{*}$, $L_{\rm [CII]}$ and SFR through a Kolmogorov-Smirnov test, it is found that the two samples are not drawn from the same distribution (p-value = 0.002). A possible explanation for this disparity is the significant scatter in $M_{*}$ and SFR values in the observed data compared to TNG50 (i.e. Figure~\ref{fig:hist_sample}).
Our results are also roughly in agreement with the expected relationship for local starburst galaxies (blue line, \citealt{2014delooze}) and for high-$z$ galaxies (red dashed line, \citealt{2018lagache}). We note that three of our simulated galaxies lie below the expected relationships, exhibiting luminosities below the observational limit. Since these galaxies represent non-detections in the observations of \citetalias{2020ginolfi}, we exclude them from any future analysis. 

Overall, the results of this analysis provide validation for the predictions made by our theoretical model on galactic scales. In the next sections we will explore what these models predict for the [CII] emission in the CGM.

\subsection{[CII] Surface brightness profiles}
\subsubsection{Averaged profiles of all gas}
\begin{figure}
    \centering
    \includegraphics[width=\columnwidth]{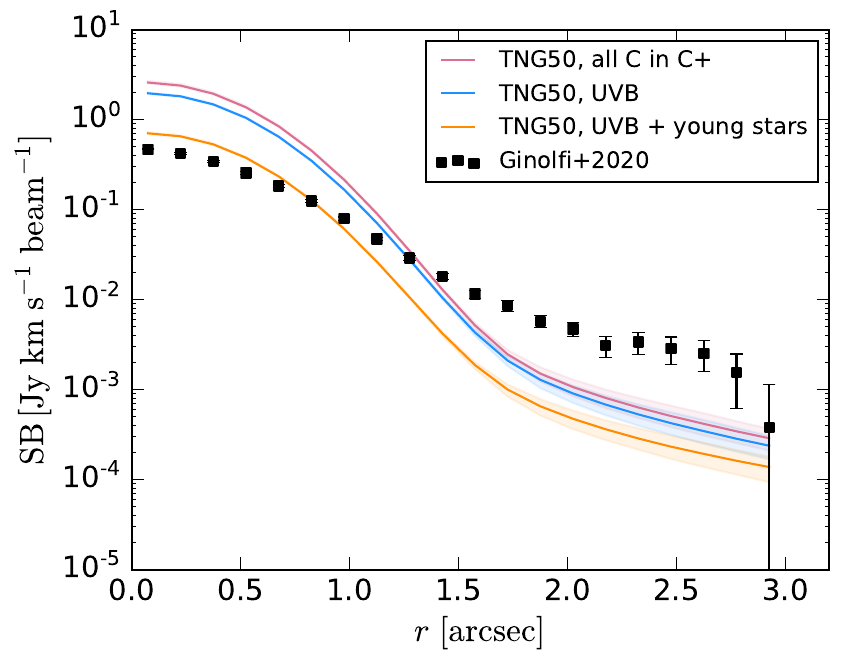}
    \caption{Averaged [CII] SB profiles obtained for our sample of 72 TNG50 galaxies at z=4, 5 and 6 (solid lines), with shaded areas enclosing a 95~percent of different results obtained by different galaxy orientations. The different line colors represent the TNG50 profiles obtained assuming C$^{+}$ abundances equal to the total C abundance (dark pink), C$^{+}$ abundance obtained considering the UVB as incident field (light blue) and C$^{+}$ abundance considering UVB and local young stars as ionization sources (orange). As a comparison, the black squares represent the results from the ALPINE survey.}
    \label{fig:sb_profiles}
\end{figure}

The study of radial [CII] surface brightness profiles in high-redshift galaxies enables a deeper understanding of the spatial distribution of [CII] emission, the properties of the cold ISM, the interplay between star formation and feedback, and the properties of the CGM (e.g., \citealt{2019fujimoto,2020fujimoto}, \citetalias{2020ginolfi}).
In Figure \ref{fig:sb_profiles} we present the resulting [CII] surface brigthness profiles of our simulated galaxies, for the three different cases of C$^{+}$ abundances calculated (represented by different colors, see legend), as explained in Section \ref{sec:cooling_rates}. The black squares show the averaged profile from \citetalias{2020ginolfi}, the solid  lines show the average (including $z$=4, 5 and 6) profiles from TNG50, and the shaded areas encompass 95~per~cent of the distribution of the averaged profiles through multiple realizations of the experiment, accounting for different orientations of the galaxies (see section \ref{sec:analysis_profiles}). The radial distances are shown up to $\sim$3~arcsec (or $\sim$19~kpc at $z$=5), which corresponds to the maximum radius analyzed in the observational profiles.

For all cases, the most central part of the profile is dominated by the PSF of the observations (up to a radius of $\sim$1.5~arcsec), while the flatter part at larger scales correspond to the inner CGM of these galaxies.
In the scenario where all carbon (C) is assumed to be converted into [CII] (dark pink line), comparing with the observations we get an overestimation of the emission at central scales ($\sim$8 times higher) and an  underestimation at scales larger than $\sim$1.5~arcsec ($\sim$6 times lower). %However, this case is the less realistic one.
When considering UVB as ionizing source (light blue line), we can see that our predicted [CII] flux is also higher ($\sim$7 times) than the observations at central scales, and lower ($\sim$7 times) at large scales. In the last case, which assumes UVB and young stars as ionizing sources (orange line), the simulated flux is comparable to the observed one at central scales (in agreement with our analysis in Section~\ref{sec:cii_sfr_relation}), but it is even lower at larger scales ($\sim$10 times lower). We note that the additional ionizing flux removes C$^{+}$ as it undergoes photo-ionization to C$^{++}$, resulting in a reduction of the central flux. Hereafter we focus on this last case which provides more realistic predictions for each galaxy emission (i.e., Figure~\ref{fig:lum_sfr}), and discuss what could cause the discrepancy on CGM scales. 
Indeed, the results of this model suggest that there is extended [CII] emission within the inner CGM. However, 
the flux normalization of these extended [CII] halos compared to the observations is not matched. In what follows we will analyze in detail the predicted SB profiles and explore the contributions from different gas phases.

\subsubsection{Contribution of star-forming gas}\label{sec:sfgas}

\begin{figure*}
    \centering
    \includegraphics[scale=0.55]{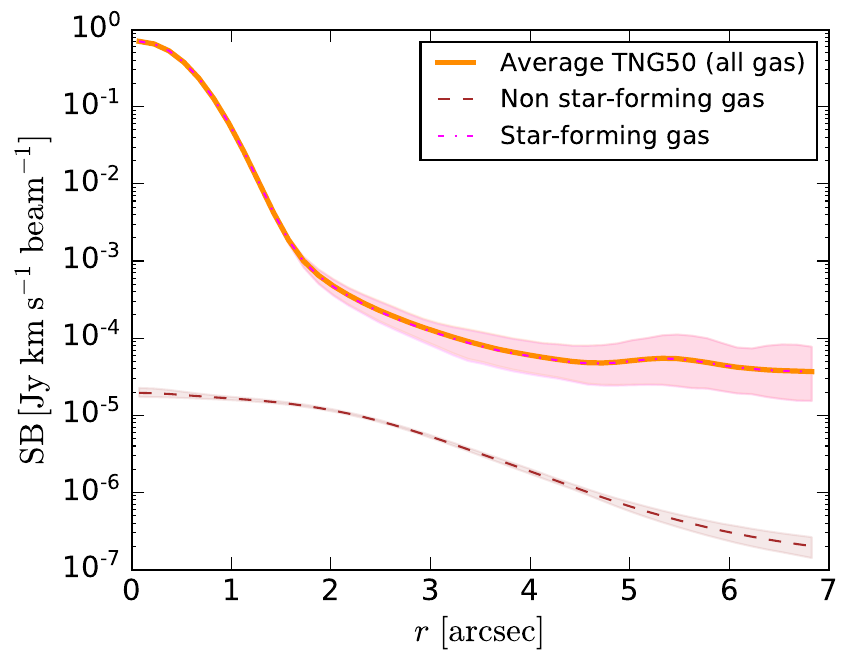}
    \includegraphics[scale=0.55]{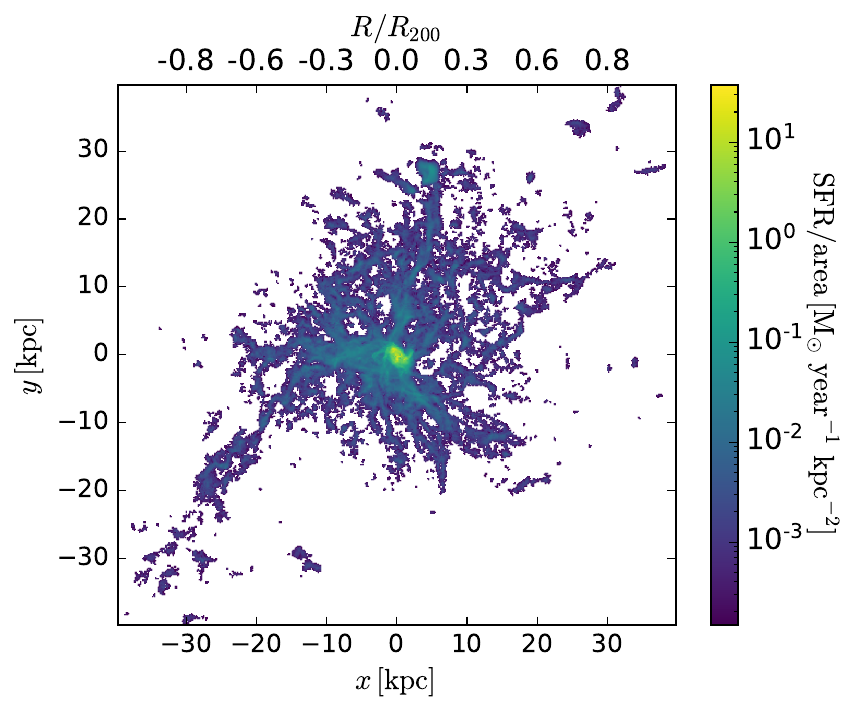}
    \caption[Left: Averaged contribution of the star-forming gas and non star-forming gas to the total {[CII]} emitting gas. Right: SFR density map for a TNG50 galaxy at $z$=5. ]{Left: Averaged contribution of the star-forming gas (magenta dot-dashed line) and non star-forming gas (brown dashed line) to the total [CII] emitting gas (orange line). The shaded areas enclose a 95~percent of different results obtained by different galaxy orientations. Right: SFR density map for a TNG50 galaxy at $z$=5. The box goes out to $\sim$40~kpc or $\sim$R$_{200}$ from the central galaxy.}
    \label{fig:profiles_sfgas}
\end{figure*}

By examining the contribution of star-forming gas (i.e., gas with $n_{\rm H}>0.13$~cm$^{-3}$, see Section \ref{sec:illustristng}) to the total averaged [CII] surface brightness profile, we can refine our understanding of the underlying physical mechanisms responsible for the observed emission.
Figure \ref{fig:profiles_sfgas} shows the total contribution of the dense star-forming gas (magenta dot-dashed line) to the total [CII] SB profile (orange line; including UVB and local young stars). The non star-forming gas contribution is shown by the brown dashed line. 
The magenta and orange lines are indistinguishable, indicating that the total [CII] emission is almost completely dominated by the star-forming gas. Since the [CII] emission in the ISM is primarily emitted from PDRs where ionized carbon atoms are heated by UV radiation from nearby young stars (e.g., \citealt{2010stacey}), this result is expected at central galactic scales (up to $\sim$1.5~arcsec), where most of the star-formation takes place. On the other hand, understanding the origin of star-forming gas dominating in the inner CGM 
(0.1$\leq R/R_{200} <$0.6, or $\sim$1.5$\leq r <$4~arcsec at $z$=5) is more complex.

Indeed, our selected TNG50 galaxies present star-forming gas extended up to R$_{200}$. This can be appreciated in Figure \ref{fig:profiles_sfgas} (right panel), that shows a projection of star-forming gas for one galaxy at $z$ = 5 (same galaxy shown in Figure \ref{fig:halo_2dflux}). In section \ref{sec:dis_sfgas} we discuss the possible origin of this large-scale star-forming gas.

\subsubsection{High and low-SFR subsamples}\label{sec:lowhighsfr}
\begin{figure}
    \centering
    \includegraphics[width=\columnwidth]{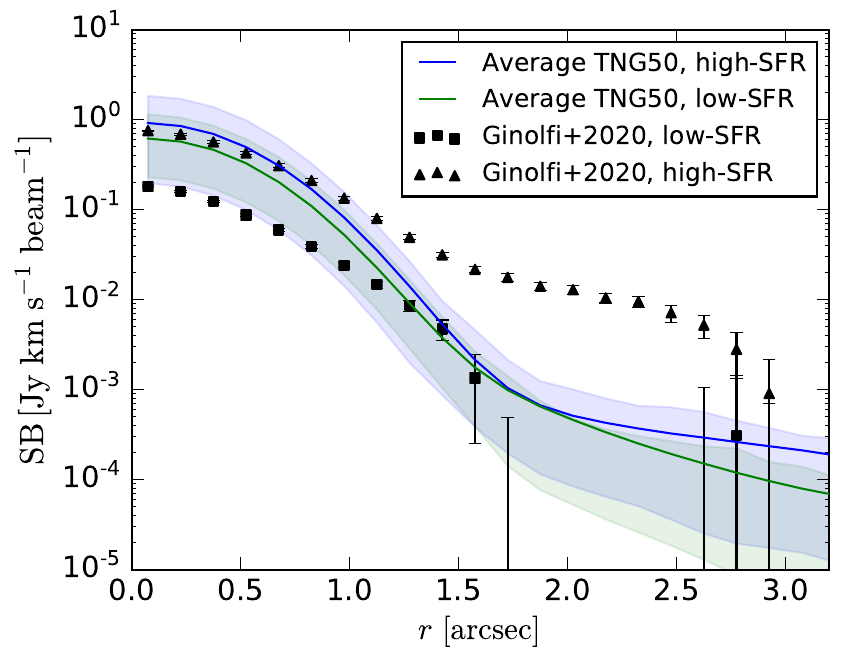}
    \caption[Averaged {[CII]} SB profiles assuming C$^{+}$ abundances from the UBV and local young stars case.]{Averaged [CII] SB profiles (solid lines) assuming C$^{+}$ abundances from the UBV and local young stars case. The blue line is the averaged profile for the high-SFR subsample, and the green one for the low-SFR subsample. The black triangles and squares represent the results for high and low-SFR galaxies of the observations, respectively. The shaded areas encompass the 2.5 and 97.5 percentiles of the distribution of all profiles (for each galaxy and each rotation).}
    \label{fig:profiles_sfr}
\end{figure}

In order to understand the relationship between the [CII] emission and SFR, \citetalias{2020ginolfi} divided their sample in two subsamples: high and low-SFR galaxies, using as threshold 25~M$_{\odot}$~year$^{-1}$. Then, to determine the typical size of the stacked [CII] line core (i.e., between [-200, 200]~km~s$^{-1}$, see Section \ref{sec:observations}), they computed SB radial profiles for both subsamples (see Figure \ref{fig:profiles_sfr}). They find that the subsample of high-SFR galaxies show a more extended profile (radial scales $>$10~kpc at $z$=5), while the profile of low-SFR galaxies drops around r$\sim$1.5~arcsec (or r$\sim$10~kpc). They claim that this result supports the scenario where star-formation feedback is responsible for the presence of [CII] halos. 

In order to further compare with the observations, we did the same analysis dividing our sample in high and low-SFR using the same threshold of 25~M$_{\odot}$~year$^{-1}$. This comparison is shown in Figure \ref{fig:profiles_sfr}, where the triangles (squares) represent the high (low) SFR subsample of \citetalias{2020ginolfi}, and the blue (green) solid curve represents the high (low) SFR subsample of the TNG50-1 galaxies. Our high and low-SFR subsamples contain 21 and 50 simulated galaxies, respectively. The shaded areas encompass the percentiles 2 and 98 of the whole distribution of averaged profiles per galaxy per rotation. The black symbols are the averaged profiles obtained by the observations.
At central galactic scales (r$<$1.5~arcsec), the profile of the high-SFR galaxies is higher ($\sim$2~times) than the one of low-SFR galaxies, which is expected (e.g., Figure~\ref{fig:lum_sfr}). Between 1.5$<$r$<$2~arcsec both profiles are in agreement. Then, at larger scales (i.e., CGM scales), the profile of the high-SFR subsample overcomes the one of low-SFR galaxies. However this is a slight difference compared to what is found in the observations, where the high-SFR galaxies profile is $\sim$4~times higher than the low-SFR galaxies profile. 
Hence, according to our model, [CII] halos are expected with similar intensities in both high and low-SFR galaxies. A possible reason for this discrepancy is that the simulations may not accurately capture the complex physical processes responsible for producing [CII] halos in galaxies, such as the effects of feedback from star formation or active galactic nuclei. A detailed discussion is presented in section \ref{sec:dis_sfgas}.

\subsubsection{Contribution of inflows and outflows}\label{flows}

Understanding the role of inflows and outflows in galaxy formation and evolution is essential to fully comprehend the physical processes that govern the growth and properties of galaxies. In this context, the identification of the [CII] contribution from each of these components allow us to disentangle the impact of inflows and outflows on the overall [CII] emission properties of the studied galaxies. In particular, cold (T$\sim$10$^{4}$K) inflows and outflows have been proposed as a possible origin of [CII] halos \citep{2019fujimoto}. 
In the case of cold gas inflows, the infalling streams can produce shock heating, resulting in the production of [CII]. Conversely, extended [CII] emission may be generated by ionized carbon, powered by outflows originating from AGN or star formation processes.
In this context, we estimated the gas radial velocities ``$v_{\rm rad}$'' with respect to the center of the halos, specifically the position at the minimum gravitational potential. We denote as outflow the gas with $v_{\rm rad}>$~0, and inflow as $v_{\rm rad}<$~0~km~s$^{-1}$.

\begin{figure}
    \centering
    \includegraphics[width=\columnwidth]{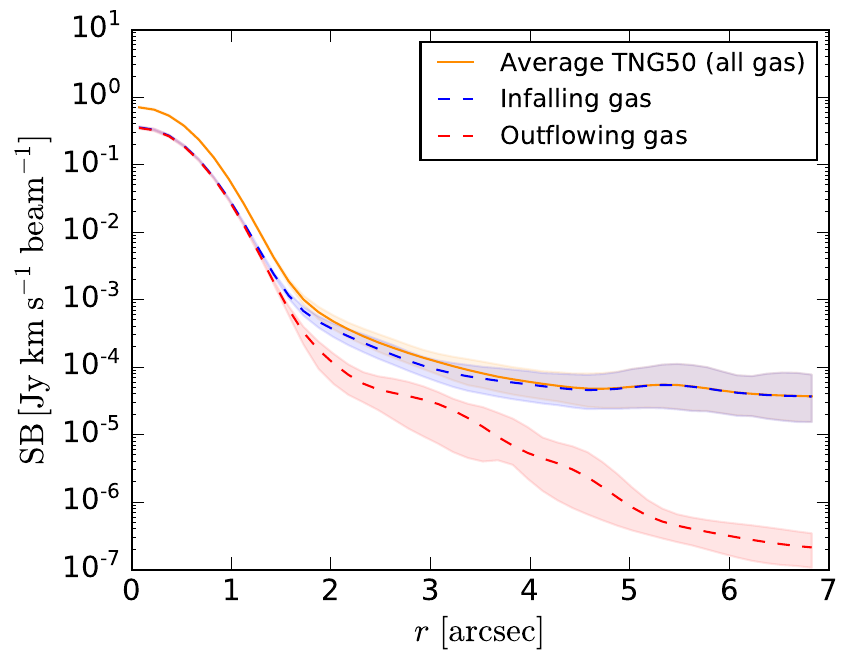}
    \caption[Averaged contribution of the infalling gas and outflowing gas to the total {[CII]} emitting gas.]{Averaged contribution of the infalling gas (blue dashed line) and outflowing gas (red dashed line) to the total [CII] emitting gas (orange line). The shaded areas represent the $2\sigma$ uncertainty obtained using 1000 samples of galaxies with randomly picked orientations, as explained in Section \ref{sec:analysis_profiles}}
    \label{fig:profiles_flows}
\end{figure}

The results of this analysis are shown in Figure \ref{fig:profiles_flows}, where we separate the contribution to the [CII] radial profile from inflowing and outflowing gas. The blue dashed line indicates the contribution of inflowing gas, while the red dashed line indicates outflows. We can appreciate that on galactic scales (up to $\sim$1.5~arcsec) the contribution of inflows and outflows is balanced. On larger scales the contribution of the infalling gas dominates, increasing with radial distance. 

\begin{figure*}
    \centering
    \includegraphics[width=\textwidth]{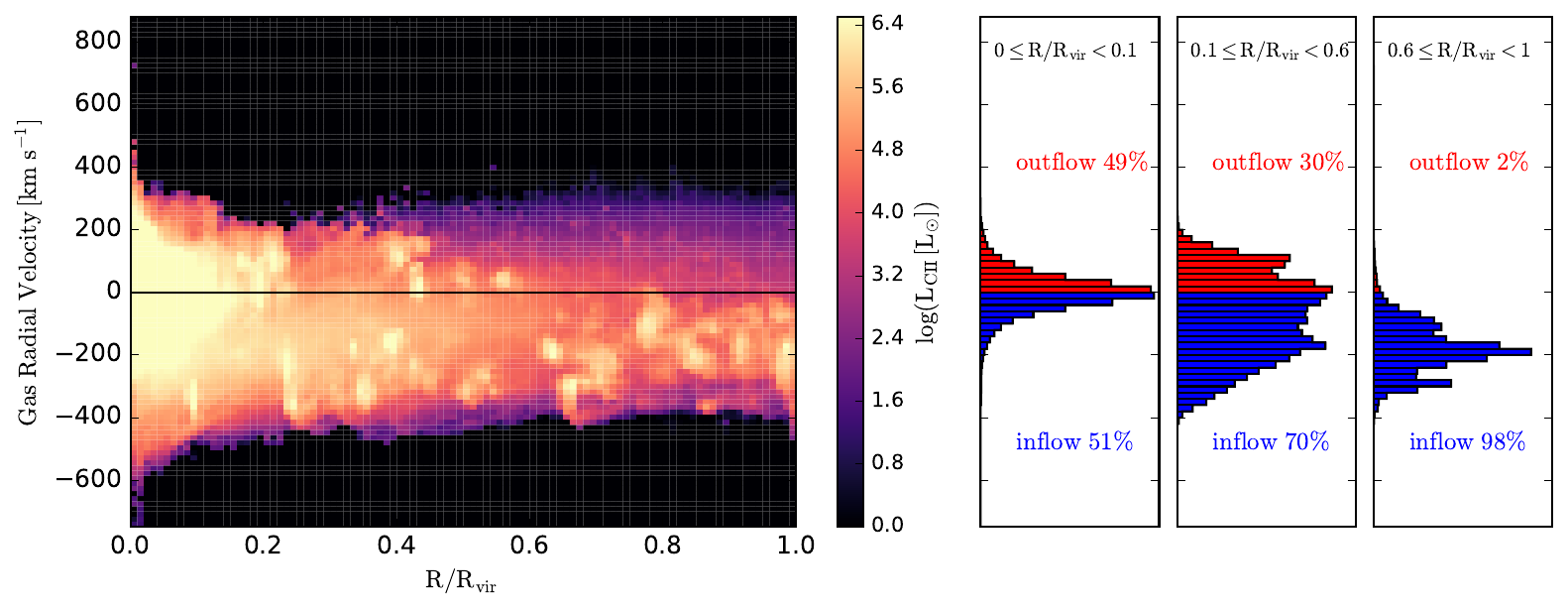}
    \caption[Left: Gas radial velocity versus radial distance, weighted by {[CII]} luminosity, for a stacking of all the selected TNG50 galaxies. Right: the histograms shows the total distribution of inflowing and outflowing gas, at different radial bins.]{Left: Gas radial velocity versus radial distance (normalized to the virial radius), weighted by [CII] luminosity, for a stacking of all the selected TNG50 galaxies (i.e, at redshifts 4, 5 and 6). Velocities greater than 0 denote outflows, and less than 0 inflows. Right: the histograms show the total distribution of inflowing and outflowing gas, at different radial bins (see legend).}
    \label{fig:2dhist_infout}
\end{figure*}

In order to further investigate [CII] as a kinematical tracer, the left panel of Figure \ref{fig:2dhist_infout} shows the gas radial velocity versus radial distance normalized by R$_{200}$, weighted by [CII] luminosity, for a stacking of all the selected halos at redshift 4, 5, 6. The histograms in the right panels show the total distribution of inflows and outflows at different radial ranges: 0$\leq R/R_{200} <$0.1, 0.1$\leq R/R_{200} <$0.6, and 0.6$\leq R/R_{200} <$1 . As a reference, the average R$_{200}$ for the TNG50 selected galaxies at $z$=5 is $\sim$40~kpc (or $\sim$6~arcsec).
For the first radial bin (inner galactic scales), our predictions indicate a balance between inflows and outflows. For the second radial bin (inner CGM), we note that the contribution of inflows exceeds the contribution of outflows by 40~per~cent. At larger scales (third radial bin), the contribution of inflows (98~\%) dominates the [CII] emission.

These results suggest that infalling gas may play a more significant role in the [CII] emission at larger scales, potentially indicating the presence of gas accretion processes that contribute to the buildup of gas reservoirs in high-redshift galaxies. These results are intriguing due to the expectation that inflowing gas from the intergalactic medium typically possesses low metallicity (e.g., \citealt{2014pallottini}). Consequently, it is anticipated that this gas would have a minimal impact on [CII] halos. Therefore, these results raise interesting implications for our understanding of metal-enriched gas inflows. 
We further explore these results in section \ref{sec:discussion}.

\subsubsection{Contribution of satellites}\label{sec:satellites}
Another potential origin of [CII] halos is related to satellite galaxies \citep[e.g.,][]{2019fujimoto}. This scenario suggests that the presence of satellite galaxies surrounding central star-forming galaxies gives rise to the observed extended [CII] emission.
In this context, we analyzed the contribution of gas originating from sources other than the central galaxy, specifically from satellite galaxies, to the total [CII] emission.
Our sample excludes systems that have ``observable'' companion galaxies, i.e. with SFR$>$1.5~M$_{\odot}$ and 
at a projected distance $<$3~arcsec from the central galaxy, to be in agreement with \citetalias{2020ginolfi}. However, we analysed the contribution of low-SFR satellite galaxies (that would not be detectable by observations) that might still be present at small and large scales. 

\begin{figure}
    \centering
    \includegraphics[width=\columnwidth]{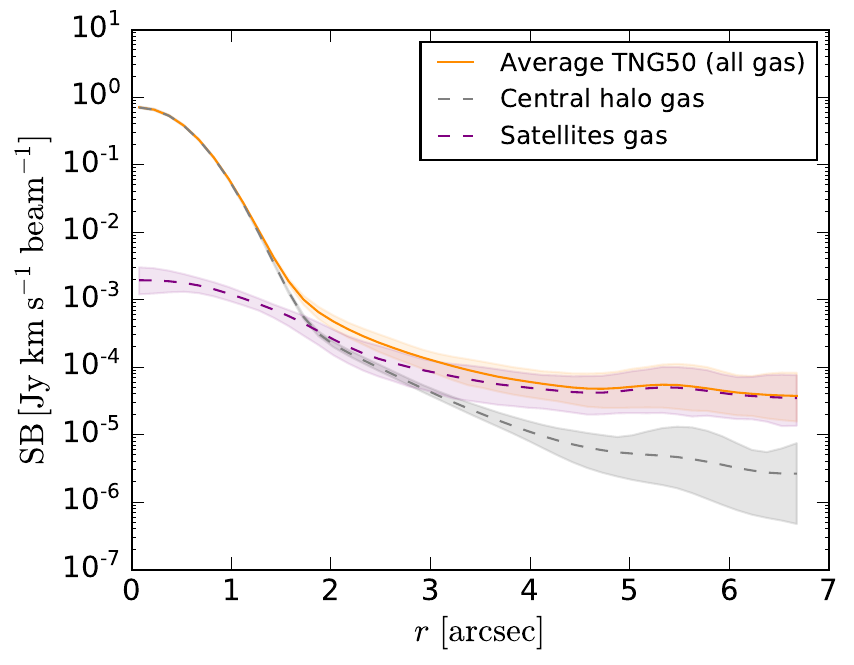}
    \caption{Averaged contribution of gas within the satellites (purple dashed line) and central galaxies (grey dashed line) to the total [CII] emitting gas (orange line). The shaded areas enclose a 95~percent of different results obtained by different galaxy orientations.}
    \label{fig:profiles_satellites}
\end{figure}

In this context, Figure \ref{fig:profiles_satellites} shows the contribution of satellites (purple dashed line) and the contribution of the central galaxies (grey dashed line) to the total [CII] profile (orange line, averaged for our sample of 72 galaxies at z=4, 5 and 6).  As expected, we note that %even though we excluded satellites on central scales, 
for the inner part of the galaxies, i.e between 0 and $\sim$2~arcsec, the contribution from the central galaxies dominate.  However, there is also a very low contribution from the satellites because of projection effects. We also note that beyond $\sim$2~arcsec the contribution of satellites starts to dominate.
This contribution increases up to $\sim$100~per~cent at R$_{200}$ ($\sim$7~arcsec).

Interestingly, these results suggest that gas that is part of or associated to satellites galaxies may play an important role in observations of [CII] halos, especially when analyzing stacked data, in contrast to what has been argued in observational works \citep[e.g.,][]{2019fujimoto,2020fujimoto}. 

\begin{figure*}
    \centering
    \includegraphics[width=\textwidth]{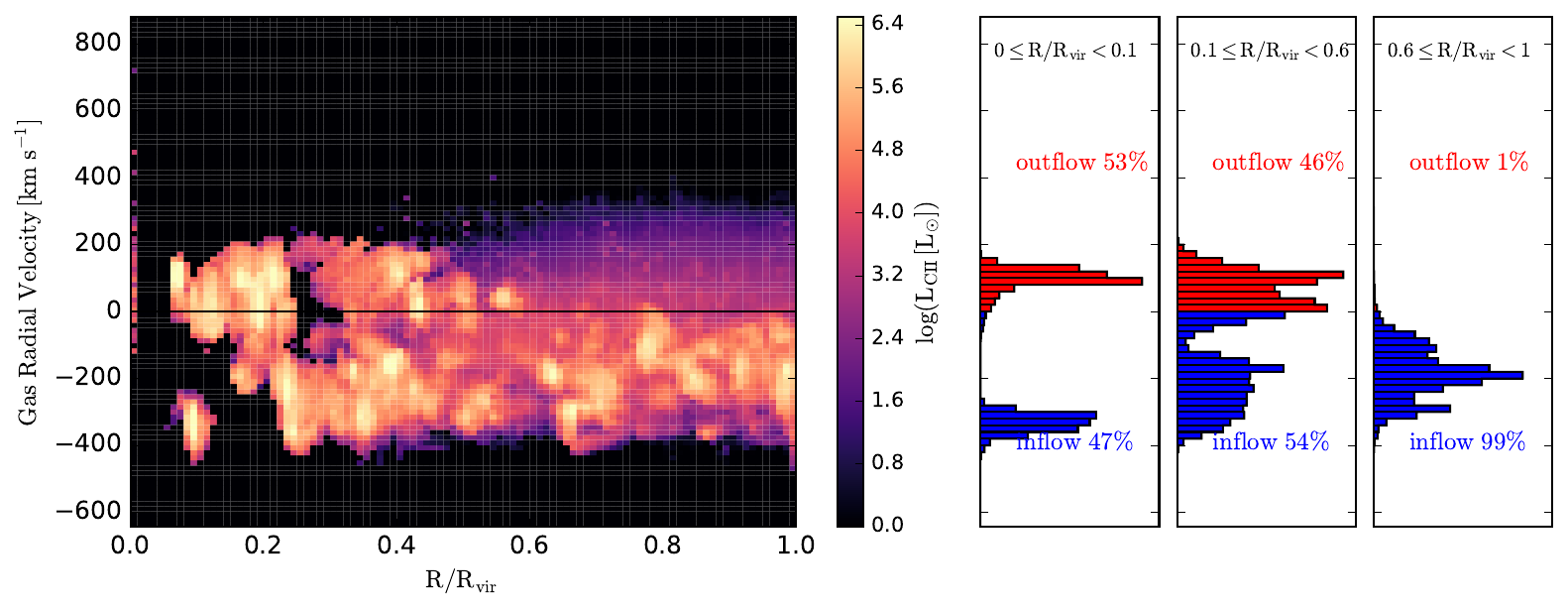}
    \caption[Same as in Figure~\ref{fig:2dhist_infout}, but for a stacking of all satellites of all the selected TNG50 halos.]{Same as in Figure~\ref{fig:2dhist_infout}, but for a stacking of all satellites (rejecting the contribution of the central galaxies) of all the selected TNG50 halos.}
    \label{fig:2dhist_infout_satellites}
\end{figure*}

To further explore the contribution of satellites, Figure~\ref{fig:2dhist_infout_satellites} shows the same analysis as Figure~\ref{fig:2dhist_infout}, but only for the gas from satellites, neglecting the contribution of the central galaxies. From the 2D histogram (left panel), it is evident that companion galaxies (clumpy structures) are already present on small scales, $\sim$0.1~$R/R_{200}$ (or 4~kpc/0.7~arcsec at $z$=5, with low contribution to the total [CII] emission). From the histograms at the right panel, we note that after 0.6~$R/R_{200}$ almost all (99~per~cent) of the gas from satellites is inflowing towards the central halo.
These results will be further elaborated and discussed in the next section.

\section{Discussion}
\label{sec:discussion}
In this section we explore different properties associated with extended [CII] emission based on the results found in this work. Subsequently, we aim to interpret these properties and discuss potential scenarios concerning the origin of [CII] halos.

\subsection{Origin of large-scale star-forming gas}
\label{sec:dis_sfgas}
The CGM is a complex and dynamic environment influenced by different physical processes operating across various scales. These processes
include gas accretion, feedback from star formation and black holes, and interactions with the surrounding environment (see \citealt{2017tumlinson} and references therein). In this context, at high-$z$ the presence of star-forming gas extending beyond the body of the central galaxies is not unexpected in the TNG model \citep[e.g.,][]{2019donnari}. In particular, \cite{2019pillepich} studied the evolution of stellar and gaseous components in star-forming TNG50 galaxies at $0<z<6$. They found that, for all masses and redshifts, the star-forming gas (as traced by H$\alpha$) settles into disky or elongated morphologies. At high-$z$, they also find that the star-forming gas often exhibits a complex structure, characterized by significant asymmetries on large scales (beyond the stellar body), which can be attributed to the prevalence of inflows and outflows during this epoch.

In Section \ref{sec:sfgas}, we show that the star-forming gas in the TNG50 galaxies is extended up to $\sim$R$_{200}$, dominating the [CII] SB profiles from inner galactic to CGM scales. One option is that this extended star-forming gas originates from the central galaxy. Galaxies undergoing active star formation can generate strong outflows that expel gas into their surroundings. This gas could then cool down and lead to star formation in the halo. Another option is that this star-forming gas comes from satellites galaxies. Interactions can trigger star formation and redistribute gas within and around galaxies, or gas can be stripped or ejected from satellite galaxies. Furthermore, the extended star-forming gas may be fueled by accretion of cold gas from the intergalactic medium along filamentary structures. 

\begin{figure}
    \centering
    \includegraphics[scale=0.57]{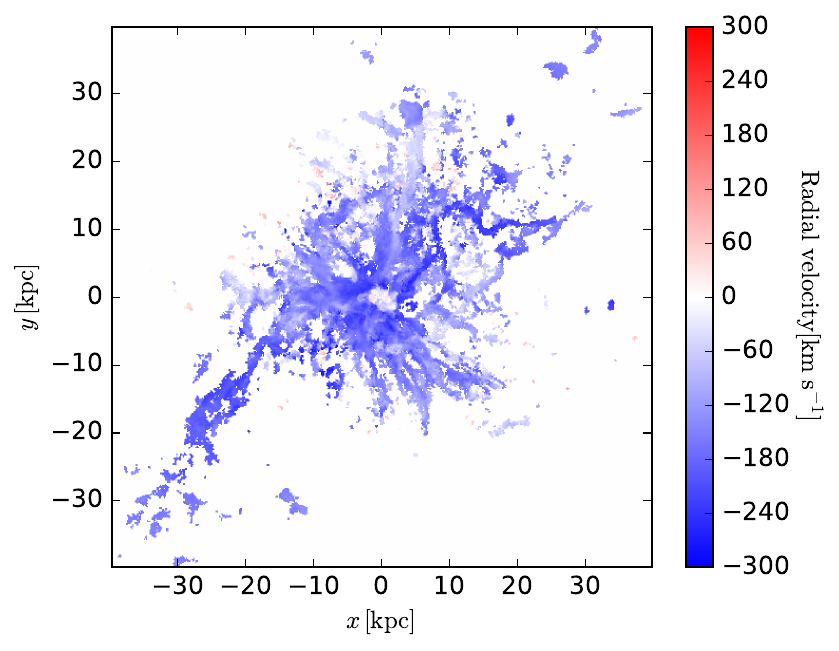}
    \caption[Average SFR-weighted radial velocity map for a galaxy at $z$=5.]{Average SFR-weighted radial velocity map for a galaxy at $z$=5. Only the star-forming gas is shown. The box goes out to $\sim$40~kpc or $\sim$R$_{200}$ from the central galaxy.}
    \label{fig:map_vradiaSF}
\end{figure}

Considering the kinematic analysis performed in Section \ref{flows} together with the contribution of satellites in Section \ref{sec:satellites}, we conclude that the large-scale (r$>$1.5~arcsec) star-forming gas is mostly infalling and belongs to the satellites galaxies. Specifically, from Figure \ref{fig:2dhist_infout_satellites} we obtain that at R/R$_{200}>$0.6 almost all the gas belonging to satellites is infalling (right third histogram).
Another illustrative example is presented in Figure \ref{fig:map_vradiaSF}, that shows an averaged SFR-weighted radial velocity map of the star-forming gas for a galaxy at $z$=5. Inflows are represented in blue and outflows in red. It is evident from the figure that the majority of the large-scale star-forming gas primarily consists of infalling gas. 

\begin{figure*}
    \centering
    \includegraphics[scale=0.48]{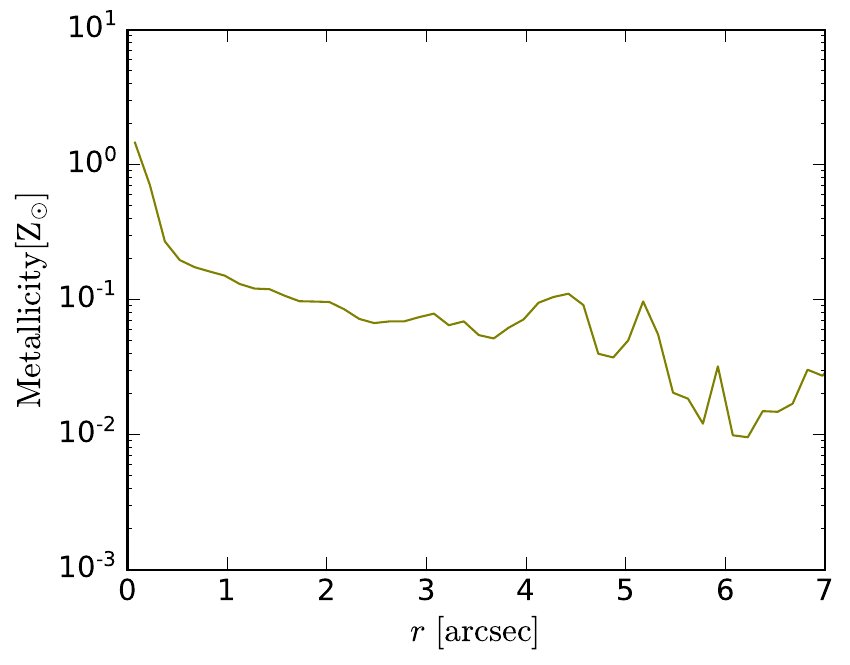}
    \includegraphics[scale=0.53]{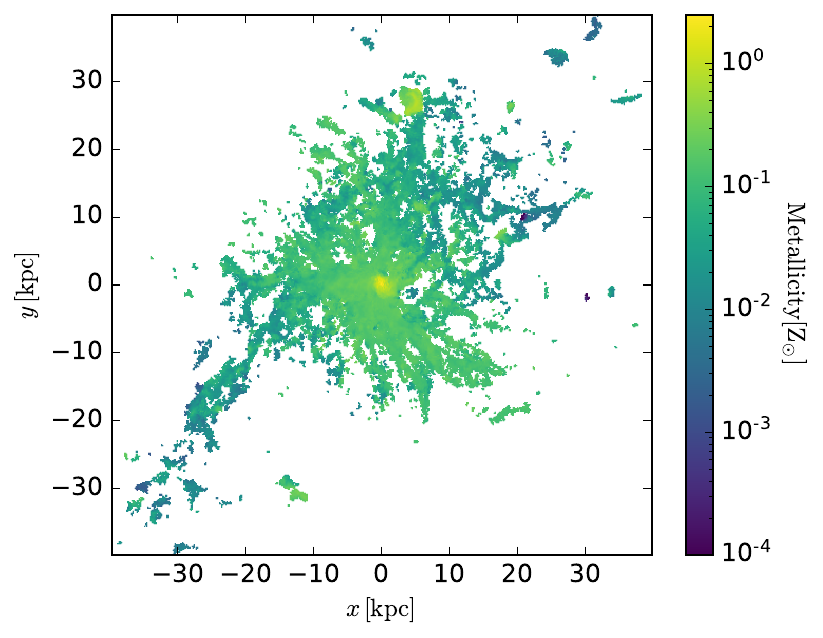}
    \caption{Left: Average radial profile of the the star-forming gas metallicity (weighted by SFR), for one galaxy at $z$=5. Right: Average SFR-weigthed metallicity map for a galaxy at $z$=5. Only the star-forming gas is shown. Its corresponding radial profile is shown in the left panel. The box size radius is $\sim$40~kpc or $\sim$R$_{200}$.}
    \label{fig:map_vradiaMET}
\end{figure*}

As a consistency check, we also analyzed the metallicity distribution of the star-forming gas. The chemical signatures within the CGM are influenced by different mechanisms such as galactic outflows from supernovae and black holes, and  the ejection or stripping of metals from satellite galaxies (e.g., \citealt{1993tegmark,2013shen,2019nelson,2019torrey,2020peroux}). Figure \ref{fig:map_vradiaMET} shows an averaged SFR-weighted metallicity map (right panel) of the star-forming gas for a galaxy at $z$=5. Its corresponding averaged radial profile is shown in the left panel. As expected, the gas that is in the central body of the galaxy exhibits higher metal enrichment (by almost 2 orders of magnitude) compared to the large-scale accreting gas. Additionally, we note that the infalling gas streams in the CGM are already metal polluted at this redshift, with typical values in the range of 10$^{-1}$-10$^{-2}$Z$_{\odot}$. As a reference, the dominant metallicity of outflowing gas for $z\sim$4 galaxies ($M_{*}\sim$10$^{9.5}$M$_{\odot}$) has been found to be $\sim$10$^{-0.7}$Z$_{\odot}$ \citep{2019nelson}, while gas from the intergalactic medium is expected to have a low-metallicty of $\sim$10$^{-3}$Z$_{\odot}$ at $z\sim$4-6 \citep{2014pallottini}.
Overall, this result is expected since metal-enriched gas is necessary for the presence of C$^{+}$ ions and consequently [CII] emission.

\subsection{Role of satellite galaxies in producing [CII] halos}

Most observational works studying [CII] halos have focused on isolated galaxies (i.e., with no observable minor or major mergers, e.g., \citealt{2019fujimoto}, \citetalias{2020ginolfi}, \citealt{2020fujimoto,2021herreracamus,2022fudamoto}). However, contamination from possible faint undetected satellites cannot be discarded. The presence of satellites may affect the gas dynamics, heating processes, or feedback mechanisms, influencing the overall [CII] emission properties. In this context, some attempts from observations have been made to estimate the contribution of such undetected satellites to the observed [CII] halos. For instance, \cite{2019fujimoto} investigated the possible contribution of satellites to the extended [CII] emission they found in a stacking of data of star-forming galaxies at $z=$5-7. They estimated radial values of the $L_{\rm CII}$/SFR ratio, finding the highest ratios in the outer regions. These values are in disagreement with those expected for local dwarf galaxies. The authors interpret this as an indication that satellites are likely not the origin of the observed [CII] halo. On the other hand, \cite{2020ginolfiB} studied one merging system at $z\sim$4.6, from the ALPINE survey. They find a [CII] halo extending up to $r\sim$15~kpc, and argue that its origin is mainly interstellar gas stripped by strong gravitational interactions, with a possibly less significant contribution from galactic ouflows and star formation from small faint satellites. \cite{2023schimek} presented a theoretical effort of simulating [CII] emission in a zoom-in simulation of a system  at $z$=6.5 undergoing a major merger. Their results indicate that the [CII] emission in the CGM is higher in an accreting filament and tidal features from the merging galaxies. Recently, \cite{2024dicesare} studied the [CII] emission in the CGM of six major mergers from the ALPINE survey. They find extended [CII] emission ($\gtrsim$20~kpc) in all the systems. By comparing with simulated galaxies, they infer that the origin of this emission is tidal stripped gas and the presence of unresolved star-forming satellites.
\begin{figure}
    \centering
    \includegraphics[width=\columnwidth]{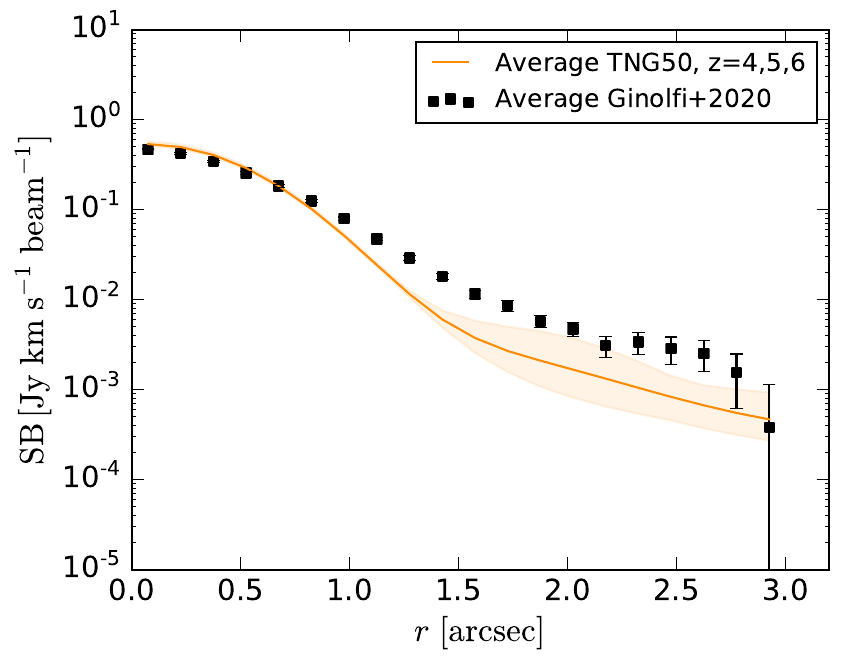}
    \caption{Averaged [CII] SB profiles obtained for a sample of 169 TNG50 galaxies at z=4, 5 and 6 (orange line) which include close satellites. The shaded area encloses 95~percent of different results obtained by different galaxy orientations. The C$^{+}$ abundances have been estimated considering UVB and local young stars as ionization sources.}
    \label{fig:profiles_parent}
\end{figure}

From the previous section and the results presented in Section~\ref{sec:satellites}, we find that star-forming gas from satellites or companion galaxies contribute substantially to the extended [CII] emission beyond $\sim$2~arcsec. %This finding could be an indication that satellites are actually the origin of the observed [CII] halos. 
This suggests that faint satellites, which may be undetected in observations, and their associated gas could be important contributors to the [CII] halos observed around galaxies. However, as we are not obtaining the full emission level of the observed [CII] profile on CGM scales, we cannot confirm that. A possibility for this missing flux is that there might be other physical processes affecting the extended [CII] emission that are not well reproduced by TNG50 (see next Section). To test what would be the maximum contribution of satellites in our model, we derived the average [CII] SB profile for a larger sample of TNG50 galaxies (169) that covers the physical properties range of in the sample of \citetalias{2020ginolfi} (see Section \ref{sec:selection_sample}), but without restricting to similar distributions in SFR and stellar mass. Importantly, we select these galaxies without removing the systems with mergers. The result is presented in Figure~\ref{fig:profiles_parent}, where the C$^{+}$ abundances were estimated by assuming UVB and young stars as ionizing sources. Interestingly, the level of [CII] emission at large scales is $\sim$8 times higher than what is found in our cleaned sample (see Figure \ref{fig:sb_profiles}), being closer to the observations. This supports the scenario in which the presence of satellites and their interactions with the central galaxies could have a significant impact on the [CII] emission in the inner CGM. This effect could be especially important in observational results based on stacking techniques.  

In summary, our model, which shows that [CII] traces star-forming gas, suggest that satellites (including faint and undetected ones) can play an important role in shaping the extended [CII] emission in the inner CGM of high-$z$ galaxies. However, further investigation is needed to fully understand the specific role of satellites in reproducing the exact levels of the observed extended [CII] emission. Even when including galaxies with close satellites, we are not able to fully match the observations. Simulations capable to resolve less massive galaxies (TNG50 resolves galaxies with M$_{*}\gtrsim$10$^{7}$~M$_{\odot}$, see Section \ref{sec:illustristng}) would be essential. Future theoretical works exploring the detailed properties of satellites, their gas content, and interactions with the central galaxies could provide insights into the mechanisms responsible for the observed [CII] halos of high-$z$ galaxies.

\subsection{Role of outflows in [CII] emission of high and low-SFR galaxies}
\label{sec:dis_highlowsfr}

Based on a spectral and cube stacking analysis, \citetalias{2020ginolfi} argue that [CII] halos trace circumgalactic gas that has been previously enriched by past outflows driven by star formation activity. They found extended [CII] emission only for the subsample of galaxies with high-SFR (SFR~$>25$~M$_{\odot}$~yr$^{-1}$), implying that galaxies with more active star formation should present more extended [CII] emission. This interpretation is further supported by semi-analytical models (\citealt{2020pizzati,2023pizzati}). For instance, \cite{2023pizzati} implemented a model to predict [CII] halos originating from outflowing gas, which incorporates key factors such as the outflow mass loading factor\footnote{$\eta = \dot{M}/\rm SFR$, with $\dot{M}$ the outflow rate.} ($\eta$), the parent galaxy SFR, and the circular velocity of the dark-matter halo. In particular, the value of $\eta$ is related to the radiative cooling of the wind and its capability to reach low temperatures (T$\sim$10$^{2}$-10$^{4}$~K). Conversely, the gravitational potential of the dark-matter halo can substantially decelerate the gas expansion. These authors compare their results with ALPINE individual galaxies (from \citealt{2020fujimoto}) finding that detected [CII] haloes are a natural outcome of starburst-driven outflows. Their model also suggests that low-mass systems also present [CII] halos, however their emission levels are too faint to be detected with the current observational capabilities.

\begin{figure}
    \centering
    \includegraphics[width=\columnwidth]{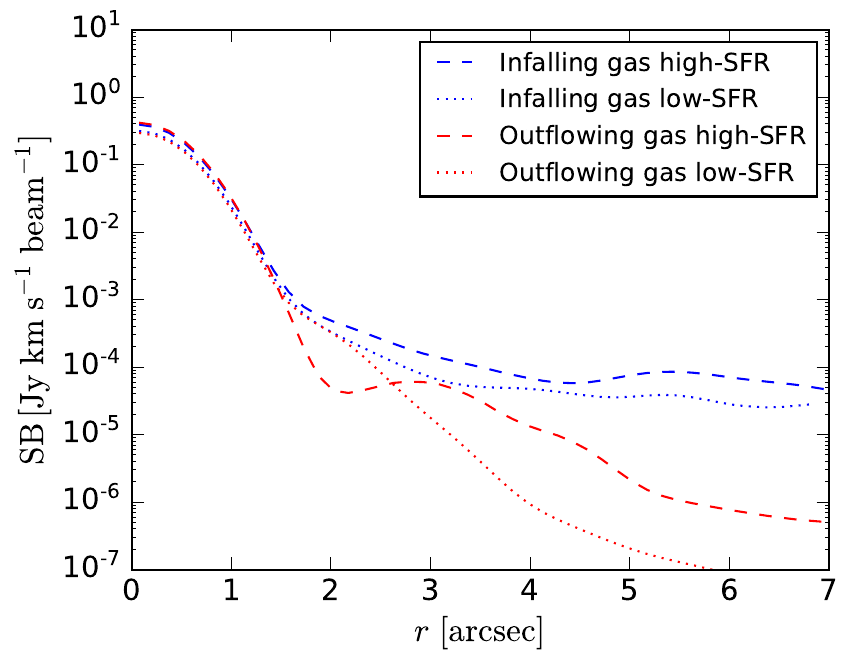}
    \caption[Averaged {[CII]} SB profiles for the outflowing (inflowing) gas only, represented by the red (blue) color.]{Averaged [CII] SB profiles for the outflowing (inflowing) gas only, represented by the red (blue) color. Our original TNG50 sample has been divided in two subsamples: high-SFR (dashed lines) and low-SFR (dotted lines). The C$^{+}$ abundances have been estimated considering UVB and local young stars as ionization sources.}
    \label{fig:profiles_flowsSFR}
\end{figure}

In Section \ref{sec:lowhighsfr}, we analyzed the [CII] SB profiles of our high-SFR and low-SFR subsamples, finding no major difference between both profiles at CGM scales, in contrast with the observational results of \citetalias{2020ginolfi}. Both subsamples present extended [CII] emission, however the emission of the high-SFR subsample should be higher (up to $\sim$20 times) to match the observations. To have more insights regarding the reason of this discrepancy, we constructed the [CII] radial SB profile of the outflowing gas only and look for differences in both SFR subsamples (for completeness, the same is done for the inflowing gas). This exercise is shown in Figure \ref{fig:profiles_flowsSFR}, where the dashed line indicates the average for high-SFR galaxies (SFR~$>25$~M$_{\odot}$~yr$^{-1}$), and the dotted line for low-SFR galaxies (SFR~$<25$~M$_{\odot}$~yr$^{-1}$). The radial axis extends to $\sim$7~arcsec ($\sim R_{200}$). We note that for galactic scales ($r<$2~arcsec), the SB of the outflowing gas is $\sim$2 times higher for the high-SFR galaxies. Then, at $r\sim$2~arcsec there is a drop in the profile of the high-SFR subsample, which then raises again, overcoming the outflowing gas profile of low-SFR galaxies at $r\sim$2.8~arcsec, until $\sim$R$_{200}$ (by a factor of $\sim$10). This indicates that outflows in the high-SFR galaxies are more energetic and hence more extended,
contributing to a larger observable [CII] SB for scales larger than $r\sim$2.8~arcsec. %at these larger scales.
However, this increased outflow contribution is still much smaller than the contribution from infalling star-forming gas in our model. If outflows were the main contributors to the [CII] halos, their phase diagram in the TNG50 simulation are not accurate enough to produce observable levels of emission. A detailed discussion about the implications of the TNG feedback model is presented in Appendix \ref{sec:appendix1}.

In summary, our predicted [CII] emission in TNG50 galaxies suggests that [CII] halos originate through a combination of contribution of star-forming gas (associated with satellites galaxies on CGM scales) and outflows from galactic scales. The contribution of inflowing gas from satellites dominates beyond $\sim$2~arcsec. However, compared to the observations we do not match the level of  [CII] emission on CGM scales. This discrepancy could be related to the limited resolution of the simulations, issues in the feedback model of TNG, and/or other physical processes not implemented in our model. The next section details the caveats of our work.

\subsection{Caveats}
\label{sec:caveats}

There are certain limitations and considerations related to the simulations and model assumptions that need to be taken into account when interpreting the results and conclusions presented in the previous sections:

\begin{itemize}
    \item {\bf Cold dense gas clumps in the CGM:} The TNG50 simulations do not fully resolve cold high-density gas within the CGM, potentially resulting in an underestimation of the [CII] emission. It is possible that the presence of cold clumps in the CGM would contribute to higher [CII] emission levels than those captured in this work. These cold clumps are known to endure within outflows, which can be intensified by cosmic rays, a factor not incorporated in IllustrisTNG (e.g., \citealt{2021rathjen}). The presence and survival of dense CGM gas is supported by observations (e.g., \citealt{2014cantalupo,2015arrigoni,2015hennawi,2021vidalgarcia}) and high-resolution ''cloud-crushing" simulations \citep[e.g.,][]{2018gronke,2020gronke,2021kanjilal}. In particular, \cite{2020nelson} attempted a resolved study of small-scale cold CGM gas in TNG50, finding the presence of cold clouds whose number and mass increase with resolution.
    
    Despite these efforts, such processes are still largely unresolved, and cosmological models of galaxy formation are not converged in their CGM properties (e.g., \citealt{2021oppenheimer}). In addition, most of the [CII] emission comes from thin PDRs (e.g., \citealt{2015gullberg}), which remain unresolved in our model. Given the critical densities associated with  C$^{+}$ and, more significantly, the dependence of $\Lambda_{\rm CII}$ on particle number density ($\Lambda_{\rm CII} \propto n^{2}$, see equations \ref{eq:lambda} and \ref{eq:rates}), it becomes necessary to incorporate higher densities, potentially through the application of density distributions on sub-grid scales (e.g., \citealt{2021olsen,2023schimek,2024khatri}).
    \item {\bf Coldest temperature:} The temperature of the star-forming gas cells is assumed to be fixed at 1000~K (cold mode, \citealt{2003springel}). While this specific value is motivated by the subgrid model used in the simulations, in reality star-forming gas is characterized by a distribution of temperatures which spans also lower values (e.g., \citealt{2009kainulainen,2011draine,2021olsen}). Colder dense gas (currently not modelled in the TNG50 suite), down to the [CII] excitation temperature of $\sim$91~K, could enhance the production of [CII] emission.
    %The simulations do not have the resolution to accurately capture the physical gas temperatures for the dense gas. As a result, in reality there might be colder gas which could enhance the production of [CII].
    \item{\bf SFRs in large-box cosmological simulations:} Reproducing the highest levels of SFR estimated in observations is a well known issue for large-box cosmological simulations (e.g., \citealt{2015furlong,2019donnari,2019pallero,2019Aoyama,2020katsianis}). The low statistics or absence of very high SFR galaxies in the simulations could affect our analysis. In particular, the number of high-SFR galaxies studied here is smaller than the number of low-SFR galaxies. Our high-SFR subsample consists of 21 galaxies (versus 50 for the low-SFR subsample), and are in the range of 25-270 M$_{\odot}$~year$^{-1}$, while in the observations they analyze 20 high-SFR galaxies in the range of 25-623~M$_{\odot}$~year$^{-1}$ (versus 20 low-SFR). The most star-forming systems in the observations, those at $\sim$600~M$_{\odot}$~year$^{-1}$, are dominating the averaged [CII] halo emission. However, in TNG50 such systems are simply not available.

    \item {\bf Turbulence dissipation in the CGM:} The CGM is expected to be a turbulent environment, influenced by different perturbations associated with galaxy interactions, accretion, and outflows (e.g., \citealt{2021vidalgarcia}). Turbulence dissipation affects the dynamics, thermal properties, and chemical enrichment of the CGM (see \citealt{2023faucher} and references therein). In the cold CGM, turbulent widths have typical values of $\sim$10-30~km~s$^{-1}$, which indicates random motions with velocity dispersions of $\sigma\sim$1000~km~s$^{-1}$ (e.g., \citealt{2018mccourt,2019rudie}). Currently, these processes are not followed by the TNG suite. Considering the above, the inclusion of sub-grid scale high-velocity turbulences could be essential to enhance the frequency of collisions between C$^{+}$ and its colliding partners, leading to an increased [CII] emission in the CGM (e.g., \citealt{2014godard}).
    \item {\bf H$_{2}$ as collisional partner:} In our post-processing model, we have neglected H$_{2}$ as colliding partner because of the reasons explained in Section \ref{sec:cooling_rates}, and especially because its physics is not modelled by the simulations. [CII] emission due to collisions with H$_{2}$ is characterized by the highest critical density when compared to collisions with e- and H (see Section \ref{sec:intro}), and such densities are not present on CGM scales in the TNG50 simulations. 
    Additionally, H$_{2}$ abundance is assumed to be dependent on metallicity (e.g., \citealt{2018krumholz}); hence, minimal H$_{2}$ is expected on halo scales where metallicities are lower than those toward galactic scales.
    %However, H$_{2}$ might be an important collisional partner out to the inner CGM. 
    However, there is growing observational evidence of the presence of molecular gas outside the main body of high-$z$ galaxies, i.e reaching the inner CGM (e.g., \citealt{2017ginolfi,2017falgarone,2019emonts,2020li,2021vidalgarcia,2023emonts}). This suggests that H$_{2}$ may indeed be an important collisional partner out to the inner CGM and contributor to the [CII] emission in those regions.
    \item {\bf Assumptions in the CLOUDY calculations:} To obtain the C$^{+}$ abundances in the simulated systems, we used \textsc{cloudy} tables by assuming a fixed escape fraction of ionising photons of 5 per cent. However, it is important to note that this escape fraction might be variable at different radial scales, as well as the optical thickness of the medium. Some works suggest that $f_{\rm esc}$ should be very small ($<$1~per~cent) to produce the observed [CII] halos (e.g., \citealt{2020pizzati,2023pizzati}), and it may also vary for different galaxy mass ranges (e.g., \citealt{2016xu}). In addition, the \textsc{cloudy} calculations and the simulations setup neglects the effects of dust. However, dust is known to be important in galaxy evolution (e.g., \citealt{2004draine}). For instance, dust is key for the chemistry of the gas by capturing selected elements onto its grains and by catalyzing formation of the H$_2$ molecule. Dust grains can contribute to gas heating mainly through the photoelectric effect (\citealt{2001WeingartnerDraine}), and gas cooling mainly through free-particle capture onto the grain surface and collision of ions with grains (e.g., \citealt{1974BurkeSilk,2004MontierGiard}). Specifically, in PDRs, where [CII] has been shown to be the main coolant, the cold dust emission spatially correlates with [CII] emission, indicating the importance of gas heating through the photoelectric effect (e.g., \citealt{2001Habart}). 
    \\
\end{itemize}

\section{Summary}
\label{sec:summary}
To investigate the nature of [CII] halos detected in observations around galaxies at $z>$4, we conducted a post-processing analysis on the Illustris-TNG50 simulation. We aimed to model [CII] emission in a sample of simulated galaxies at $z\sim$4-6. Specifically, by incorporating C$^{+}$ abundances derived by assuming UV background and young stars as radiation sources, we estimate [CII] emissivities and generate mock observations, comparing our results with the ALPINE [CII] emitting galaxies investigated by \citetalias{2020ginolfi}. This approach allowed us to study the possible origin of extended [CII] emission in high-$z$ galaxies, and discuss the contribution of different factors to this emission.
The main results of this work are summarized as follows:

i) Our model reproduces the $L_{\rm CII}$ versus SFR relationship found by previous works in local and high-$z$ galaxies. This result provides validation for the predictions from our model on galactic scales.

ii) We derived an averaged [CII] SB profile and compared it to the observed data. Our predicted [CII] emission levels on central scales ($r<$1.5~arcsec or $\sim$10~kpc at $z$=5) are in agreement with the observations. However, at CGM scales our results underestimate the [CII] emission levels by a factor of $\sim$10.

iii) The [CII] emission in our systems is dominated by the dense star-forming gas, as expected. This gas extends up to the virial radius.

iv) From a kinematic analysis, we find that on inner galactic scales ($r<$0.7~arcsec or 5~kpc at $z$=5) the [CII] emission traces a balance between inflows and outflows (contribution of 51 and 49~per cent, respectively).

v) The contribution of infalling gas from satellites dominates the [CII] emission in the CGM, specifically after $r\sim$2~arcsec (or $\sim$13~kpc at $z$=5). This suggests that faint and undetected satellites and their associated gas may play a significant role in shaping the observed extended [CII] emission.

vi) We find further tension with the observations when comparing the SB profiles for the low and high-SFR galaxies, finding no significant  difference in the extended [CII] emission levels (up to r$\sim$3~arcsec or $\sim$20~kpc at $z$=5) between the two subsamples. We propose that this could be an effect of the feedback model and the limited resolution of the TNG simulations.

vii) According to our results, the origin of [CII] halos is a combination of contribution of gas from satellites and outflows from the central galaxies. However, the contribution of infalling gas by galaxies completely dominates at 4$<r<$7~arcsec (0.6$<r/R_{200}<$1). \\

This theoretical effort  highlights the complex nature of extended [CII] emission in the inner CGM of high-redshift galaxies. Further investigations based on state-of-art simulations and observations are necessary to fully understand the specific role of satellites and other physical processes (e.g., galactic winds, turbulences) in reproducing the observed [CII] halos. In particular, cosmological simulations with higher resolution in the ISM and CGM are needed. While running cosmological simulations with higher refinements is currently computationally expensive and requires tuning, one could achieve improved results with the use of more detailed subgrid prescriptions based on radiative transfer calculations run on high-resolution turbulent boxes (Buhlmann et al. in prep). On the other hand, future observations from large sub-mm single-dish telescopes (e.g., AtLAST; \citealt{2020ATLAST}) in combination with high-resolution interferometric observations (e.g., CRISTAL; \citealt{2024Posses}) will be able to improve our general knowledge of the CGM and specifically of [CII] halos in high-$z$ objects and verify whether there are any bias in the current small observational samples (e.g., selection bias favoring more luminous [CII] halos at a given SFR). 
%The results of the observations may be affected by a selection bias favoring more luminous [CII] halos at a given SFR..

\begin{acknowledgements}
The authors thank the anonymous referee for their detailed comments that helped improving the manuscript.
The authors like to thank Dylan Nelson for insightful comments after the presentation of an earlier version of this work.
A.O. has been funded by the \emph{Deut\-sche For\-schungs\-ge\-mein\-schaft} (DFG, German Research Foundation) – 443044596.
    
\end{acknowledgements}

\bibliographystyle{aa} % style aa.bst
\bibliography{paper.bib} % your references Yourfile.bib
\begin{appendix}
\section{TNG feedback model and [CII] emission}
\label{sec:appendix1}
The TNG model assumes that star formation feedback drives galactic outflows, which are launched from star-forming gas with a wind velocity ($v_{\rm w}$) that depends on the local dark matter velocity dispersion ($\sigma_{\rm DM}$, e.g., \citealt{2013vogelsberger}). The outflow mass loading is determined by the supernova energy and the assumed wind speed, while the metal content is a fraction of ISM metallicity. The model uses a kinetic wind scheme where wind particles are created and decoupled from the gas until they leave the dense ISM. Outside the dense medium, the wind particles recouple with the gas, transferring their mass, momentum, metals, and thermal energy to the surrounding medium. Wind particles re-couple to the gas cell when: 1) the density drops below a threshold ($n_{\rm H}$ = 0.05~$\times$~density threshold for star formation = 0.006~cm$^{-3}$), or 2) after a maximum travel time has been reached (0.025~$\times$~Hubble time at respective $z$) \citep{2018pillepich}. 

For the TNG50 galaxies at $z$=5 (median of the sample) studied here, the average distance\footnote{To estimate this distance, an average $n_{\rm H}$ profile has been computed, considering 3D radial bins of $\sim$0.4~arcsec. This profile then provided the average distance at which the density threshold is reached. } at which the wind particles would re-couple with the gas from the condition 1), is $\sim$5~arcsec (or $\sim$32~kpc) and $\sim$4.6~arcsec (or $\sim$30~kpc) for the high and low-SFR subsamples, respectively. On the other hand, following condition 2), the maximum distance\footnote{This distance is given by $v_{\rm w} \times \text{current Hubble time} \times 0.025$. We first estimated $v_{\rm w}$ that depends on $\sigma_{\rm DM}$ (see equation 1 in \citealt{2018pillepich}). To estimate $\sigma_{\rm DM}$, we assumed that $V_{\rm max}\sim$1.45$\sigma_{\rm DM}$ \citep{2010okamoto}, with $V_{\rm max}$ the maximum circular velocity of the host dark matter halo. By assuming a NFW profile, $V_{\rm max}$ was estimated as $V_{\rm max}$ = $\sqrt{0.2162 c/f(c)} V_{\text{vir}}$, with $V_{\text{vir}}$ the virial halo velocity, $c$ the concentration parameter, and $f(c) = ln(1+c) - c /(1+c)$ (e.g., \citealt{2014dutton}).} after the maximum travel time is $\sim$2.8~arcsec (or $\sim$18~kpc) and $\sim$2.4~arcsec (or $\sim$15~kpc) for the high and low-SFR subsamples, respectively. These results indicate that for the case of our selected halos, the wind particles re-couple with the gas cells according to the second criteria, which is first met. The different radial distances estimated for both subsamples confirm that the outflows of the high-SFR galaxies propagate to larger distances compared to the low-SFR galaxies (Figure \ref{fig:profiles_flowsSFR}). 
On the other hand, the distance of $\sim$18~kpc ($\sim$2.8~arcsec) obtained for the high-SFR subsample is comparable to the distance that is reached by the outflows in the observations of \citetalias{2020ginolfi} ($r\sim$15~kpc).  It also coincides with the upturn of the outflowing gas SB profile, ocurring  immediately after the drop in this profile ($\sim$2~arcsec, Figure \ref{fig:profiles_flowsSFR}). This suggests that, as the wind particles re-couple with the gas and transfer their properties, it can lead to a change in the gas properties and potentially in an increase in the [CII] emission. Therefore, if all of the observed extended [CII] emission is attributed to stellar winds, it would be necessary to incorporate a new subgrid prescription able to enhance the wind contribution (i.e., the outflow mass loading factor) significantly. This increase could be achieved by including a prescription for an entrained clumpy, cold dense phase that is currently missing (e.g., \citealt{2013hamann,2016liang,2016faucher,2018mccourt,2018gronke,2020gronke}).

Finally, it is important to discuss the effect of AGN feedback. While the observed samples do not report the evidence for AGN activity in the targeted galaxies down to current observational limits, all our selected TNG50 galaxies have AGNs (L$_{\rm BOL}\lesssim$10$^{45.5}$~erg~s$^{-1}$) affecting the surrounding gas with thermal mode feedback 
(i.e., high accretion rates, average $\dot{M}/\dot{M_{\rm Edd}}$=~0.5). The average AGN feedback energy is higher for the high star-forming sample compared to the low star-forming sample ($\Delta E\, = 10^{43.9}\mathrm{erg\ s^{-1}}$ and $10^{43.3}\mathrm{erg\ s^{-1}}$, respectively). This stronger AGN feedback in the most star-forming galaxies may lead to higher gas temperatures, %(right panel of Figure~3.17), %and ionization levels, 
resulting in similar or lower [CII] emission compared to the low star-forming galaxies. Importantly, an AGN would also affect the $C^{+}$ abundances with its hard radiation. This effect is not taken into account in our calculations, but would increase the tension with observations as we expect Carbon to be more ionized when accounting for an AGN contribution.

\section{Example of input \textsc{cloudy} file}
\label{sec:appendix2}

In the current version of \textsc{cloudy} (C17.03), the UVB models of \citet{2019khaire}
can be easily accessed with the \verb|table KS18 redshift [value]|, while for our specific choice of young stellar radiation field we pass the spectrum with the \verb|table SED "[name_spectral_file]"| and provide the normalization as the logarithm of the ionizing photon flux \verb|phi(H) [value]| corresponding to each radiation field normalization $\phi$. Lines starting with "\#" in the following input file example denote comments. The output file *.C\_ionf contains the ionization fractions for all the levels of the carbon atom.

\begin{verbatim}
# set the hydrogen number density
constant density
hden [value] log
# set the temperature
constant temperature, t = [value] K log
# set the metallicity
metals [value] log
# set the incident radiation fields 
table KS18 redshift [value]
CMB redshift [value]
table SED "[name_spectral_file]"
phi(H) [value]
# set the stopping criteria
stop zone 1
# set the output
set save prefix "[name_input_file]"
save element carbon last ".C_ionf"
iterate to convergence
\end{verbatim}

For the UVB-only models, we use the same input file, but remove the two lines \verb|table SED "[name_spectral_file]"| and \verb|phi(H) [value]|.

\end{appendix}

\end{document}